\let\TeXyear\year
\documentclass{ieeeaccess}

\let\year\TeXyear

\ifCLASSINFOpdf
\else
\fi
\usepackage{url}

\usepackage{cite}
\usepackage{amsmath,amssymb,amsfonts}
\usepackage{algorithmic}
\usepackage{graphicx}
\graphicspath{{./graphics/}}

\usepackage{setspace}   
\usepackage[font={sf,scriptsize,stretch=0.84},
labelfont={bf,color=accessblue}
]{caption}  

\usepackage{textcomp}
\usepackage[dvipsnames]{xcolor}
\usepackage{siunitx}
\usepackage[nolist]{acronym}
\usepackage{booktabs}
\usepackage{pstricks}
\usepackage{pgfplots}
\pgfplotsset{compat=newest}
\usetikzlibrary{plotmarks}
\usetikzlibrary{arrows.meta}
\usepgfplotslibrary{patchplots}
\usepackage{tikz}
\usepackage{tikzscale}
  \NewSpotColorSpace{PANTONE}
  \AddSpotColor{PANTONE} {PANTONE3015C} {PANTONE\SpotSpace 3015\SpotSpace C} {1 0.3 0 0.2}
  \SetPageColorSpace{PANTONE}%
\pgfplotsset{plot coordinates/math parser=false}
\usepackage{grffile}
\usetikzlibrary{external}
\usepackage{cleveref}
\Crefname{equation}{Equation}{Equations}
\Crefname{figure}{Figure}{Figures}
\Crefname{tabular}{Table}{Tables}
\crefname{equation}{}{}
\crefname{figure}{Fig.}{Figs.}
\crefname{tabular}{Tab.}{Tabs.}
\usepackage{pdfpages}

\newcommand{\ie}{i.e.}
\newcommand{\eg}{e.g.}

\newcommand{\numTones}{Q}
\newcommand{\indTones}{q}
\newcommand{\numDelay}{M}
\newcommand{\indDelay}{m}

\newcommand{\eye}[1]{\vec{I}_{#1}}

\newcommand{\diffFreq}{\Delta f}
\newcommand{\diffd}{\,d}
\newcommand{\numBS}{A}
\newcommand{\indBS}{a}
\newcommand{\numMS}{K}
\newcommand{\indMS}{k}

\renewcommand\vec{\mathbf}

\newcommand{\depsMSDelay}{_{\indMS,\indDelay}}
\newcommand{\depsMSpDelay}{_{\indMS',\indDelay}}

\newcommand{\depsDelay}{_{\indDelay}}
\newcommand{\RXsinglesymb}{y}
\newcommand{\RXsymb}{\vec{\RXsinglesymb}}
\newcommand{\TXsinglesymb}{x}
\newcommand{\TXsymbNoDeps}{\vec{x}}

\newcommand{\NsinglesymbNoDeps}[1]{z{}}
\newcommand{\Nsinglesymb}{z}

\newcommand{\Nsymb}{\vec{\Nsinglesymb}}

\newcommand{\HsymbNoDeps}{\vec{H}}
\newcommand{\Hsymb}[1]{\HsymbNoDeps{}_{#1}}
\newcommand{\HsinglecoefNoDeps}[1]{h}

\newcommand{\HsinglesymbNoDeps}[1]{\vec{h}}
\newcommand{\Hsinglesymb}[1]{\HsinglesymbNoDeps{}_{#1}}
\newcommand{\GsinglesymbNoDeps}[1]{\vec{g}}

\newcommand{\BFNoDeps}{\vec{W}}
\newcommand{\BF}[1]{\BFNoDeps_{#1}}
\newcommand{\BFvecNoDeps}{\vec{w}}
\newcommand{\BFvec}[1]{\BFvecNoDeps{}_{#1}}

\newcommand{\Herm}{^{\mathrm{H}}}
\newcommand{\Transp}{^{\mathrm{T}}}

\newcommand{\DistAsComplexGauss}[1]{\sim\mathcal{CN}\left(#1\right)}

\DeclareMathOperator{\SINR}{SINR}

\newcommand{\reg}[1]{$\mathrm{R}_{#1}$}
\newcommand{\mreg}[1]{\mathrm{R}_{#1}}
\newcommand{\BSconf}{\ac{BS} conf.\@}

\newcommand{\TBD}[1]{}
\newcommand{\TBC}[1]{}
\newcommand{\TBW}[1]{}
\newcommand{\oldnew}[2]{#2}

\DeclareSIUnit \belm {Bm}

\begin{document}

\history{{Received 11 August 2022, accepted 16 August 2022, date of publication 19 August 2022, date of current version 29 August 2022.}}
\doi{{10.1109/ACCESS.2022.3200365}}

\title{Towards Cell-Free Massive MIMO: A Measurement-Based Analysis}
\author{\uppercase{David Löschenbrand}, \IEEEmembership{Member, IEEE},
        \uppercase{Markus Hofer}, \IEEEmembership{Member, IEEE}, 
        \uppercase{Laura Bernadó}, \IEEEmembership{Member, IEEE},
        \uppercase{Stefan Zelenbaba}, \IEEEmembership{Member, IEEE}, and 
        \uppercase{Thomas Zemen}, \IEEEmembership{Senior Member, IEEE}
        }
\address[]{AIT Austrian Institute of Technology GmbH, Giefinggasse 4, 1210 Vienna, Austria (e-mail: firstname.lastname@ait.ac.at)}
\tfootnote{This paper is a result of the project DEDICATE (dedicate.ait.ac.at). The DEDICATE project is funded within the Principal Scientist grant “Dependable Wireless 6G Communication Systems” at the AIT Austrian Institute of Technology. 
}

\markboth
{D. Löschenbrand \headeretal: Towards Cell-Free Massive MIMO: A Measurement-Based Analysis}
{D. Löschenbrand \headeretal: Towards Cell-Free Massive MIMO: A Measurement-Based Analysis}
%

\corresp{Corresponding author: David Löschenbrand (e-mail: david.loeschenbrand@ait.ac.at).}
\begin{abstract}
Cell-free widely distributed massive \ac{MIMO} systems utilize radio units spread out over a large geographical area. The radio signal of a \ac{UE} is coherently detected by a subset of \acp{RU} in the vicinity of the \ac{UE} and processed jointly at the nearest \ac{BPU}. This architecture promises two orders of magnitude less transmit power, spatial focusing at the \ac{UE} position for high reliability, and consistent throughput over the coverage area. All these properties have been \oldnew{investigates}{investigated} so far from a theoretical point of view. To the best of our knowledge, this work presents the first empirical radio wave propagation measurements in the form of time-variant channel transfer functions for a linear, widely distributed antenna array with 32 single antenna \acp{RU} spread out over a range of \SI{46.5}{\meter}. The large aperture allows for valuable insights into the propagation characteristics of cell-free systems. Three different co-located and widely distributed \ac{RU} configurations and their properties in an urban environment are analyzed in terms of time-variant delay-spread, Doppler spread, path-loss, and the correlation of the local scattering function over space. For the development of 6G cell-free massive \ac{MIMO} transceiver algorithms, we analyze properties such as channel hardening, channel aging as well as the \ac{SINR}. Our empirical evidence supports the promising claims for widely distributed cell-free systems.
\end{abstract}


\titlepgskip=-15pt

\maketitle

%

\begin{acronym}
	\acro{ACK}{acknowledgement}
	\acro{AIT}{Austrian Institute of Technology GmbH}
	\acro{BPU}{baseband processing unit}
	\acro{BS}{base station}
	\acro{CDF}{cumulative distribution function}
	\acro{CIR}{channel impulse response}
	\acro{CP}{cyclic prefix}	
	\acro{CSI}{channel state information}
	\acro{CTF}{channel transfer function}
	\acro{DL}{downlink}
	\acro{DPS}{discrete prolate spheroidal}
	\acro{DSD}{Doppler spectral density}
	\acro{DSFT}{discrete symplectic Fourier transform}
	\acro{DU}{distributed unit}
	\acro{eMBB}{enhanced mobile broadband}
	\acro{GIS}{geographic information system}
	\acro{GNSS}{global navigation satellite system}
	\acro{GoF}{goodness of fit}
	\acro{i.i.d.}{independent and identically distributed}
	\acro{KS}{Kolmogorov-Smirnov}
	\acro{LOS}{line of sight}
	\acro{LS}{least squares}
	\acro{LSF}{local scattering function}
	\acro{LTE}{long term evolution}
	\acro{MIMO}{multiple-input multiple-output}
	\acro{MMSE}{miminum mean square error}
	\acro{MRT}{maximum ratio transmission}
	\acro{MS}{mobile station}
	\acro{MRC}{maximum ration combining}
	\acro{NLOS}{non line of sight}
	\acro{OFDM}{orthogonal frequency-division multiplexing}
	\acro{OP}{orthogonal precoding}
	\acro{PA}{power amplifier}
	\acro{PDP}{power delay profile}
	\acro{PPS}{pulse per second}
	\acro{PDF}{probability density function}
	\acro{RF}{radio frequency}
	\acro{RHH}{remote radio head}
	\acro{RMS}{root mean square}
	\acro{RT}{ray tracing}
	\acro{RU}{radio unit}
	\acro{RZF}{regularized zero-forcing}	
	\acro{RX}{receiver}
	\acro{SE}{spectral efficiency}
	\acro{SDR}{software-defined radio}
	\acro{SINR}{signal to interference and noise ratio}
	\acro{SNR}{signal to noise ratio}
	\acro{SRS}{sounding reference signal}
	\acro{SVS}{singular value spread}
	\acro{TDD}{time division duplex}
	\acro{TX}{transmitter}
	\acro{UE}{user equipment}
	\acro{UL}{uplink}
	\acro{UPS}{uninterruptible power supply}
	\acro{URLLC}{ultra-reliable low latency communication}
	\acro{USRP}{universal software radio peripheral}
	\acro{UTD}{Uniform Theory of Diffraction}
	\acro{V2I}{vehicle to infrastructure}
	\acro{WAA}{wide aperture array}
	\acro{WR}{White Rabbit}
	\acro{ZF}{zero-forcing}
\end{acronym}
\acresetall 
\setcounter{page}{89232}
\section{Introduction}
\label{sec:Introduction}

\IEEEPARstart{C}{ell-free} massive \ac{MIMO} systems are a revolutionary new architecture for future mobile communications systems \cite{Demir21}. \Acp{RU} are distributed in space over a large geographical area and processed coherently at a central \ac{BPU}. This architecture enables similar properties as known from massive MIMO systems, such as high spectral efficiency using linear processing, and exploiting channel reciprocity between uplink and downlink. However, {crucial improvements are achieved by distributing the \acp{RU} in space}, such as (i) a strong transmit energy reduction due to reduced distance to the \ac{UE}, (ii) a consistent throughput over the coverage area avoiding the strong throughput drop at cell edges, and (iii) mitigation of large-scale fading \cite{zhang2019cell}.

So far, cell-free massive \ac{MIMO} systems have been explored from a theoretical point of view (see \cite{Demir21} and the references therein). Empirical evidence on propagation conditions in cell-free systems is missing, although {being} of fundamental importance. Cell-free systems can be seen as a distributed antenna array with a large aperture, causing common far-field assumptions (\eg{,} plane wave propagation) to not hold and propagation conditions to differ substantially from \ac{RU} to \ac{RU}. In other words, the propagation conditions for mobile users are non-stationary in space and time.

In this work, we use a \ac{SDR} based measurement system \oldnew{for characterizing}{to characterize} the propagation conditions in cell-free massive \ac{MIMO} systems with mobile users. A distributed massive \ac{MIMO} testbed was established at AIT Austrian Institute of Technology \cite{Loschenbrand2019}. The architecture of \cite{Loschenbrand2019} allows {for} fully parallel radio channel measurements \oldnew{for}{of} vehicular users as well as data transmission including channel prediction \cite{Loeschenbrand2020}. For widely distributed and cell-free radio channel measurements, the architecture of \cite{Loschenbrand2019} is extended, enabling a distribution of antenna elements over up to \SI{90}{\meter}. This is achieved by separating the down-conversion from \ac{RF} to baseband into a \ac{DU} and performing baseband processing in a \ac{BPU}. Phase coherent operation is achieved by a dedicated synchronization network using a \SI{10}{\mega\hertz} clock and a one pulse per second signal with \oldnew{dedicated}{suitable} distribution circuits.

\subsection{Related work and literature overview}
The research work on cell-free massive \ac{MIMO} has increased tremendously over the last five years, see, \eg{,} \cite{obakhena2021application, zhang2019cell, interdonato2019ubiquitous, Demir21} and the references therein, as it is considered an enabling technology for beyond 5G systems. While anticipated key characteristics are already known through extensive analysis and simulation, empirical validation is \oldnew{largely missing}{missing in most instances}. 

{In \cite{interdonato2019ubiquitous}, the authors put forward a cost-efficient architecture called \textit{radio stripes} for cell-free systems. The architecture includes antennas and associated processing units, synchronization, data transfer, and power supply within a single cable, thus facilitating deployment considerably. However, no empirical data obtained with this approach is publicly available. With the proposed measurement framework presented in this paper, linear deployment strategies similar to the radio stripes technology can be analyzed based on real-world data for the first time.}

Wireless propagation channel measurements \oldnew{with a virtual wide-aperture array at \SI{2.6}{\giga\hertz} with an aperture size of \SI{7.3}{\meter}}{at \SI{2.6}{\giga\hertz} with a virtual wide-aperture array measuring \SI{7.3}{\meter}} were first reported in \cite{payami2012channel}. The proposed measurement approach involves one antenna that is moved to form a virtual array, and is only suitable for static scenarios without mobility.

In \cite{Martinez2014} the authors describe a comparative study of \ac{MIMO} antenna geometries, with aperture sizes ranging from \SI{0.3}{\meter} to \SI{6}{\meter}. Eight fully parallel receive units and fast switching was used to characterize the system with 64 \ac{BS} antenna elements.

{A first cell-free channel measurement using a drone with a single transmit antenna that flies from one \ac{RU} position to the next was reported in \cite{Choi21}.} This method does not provide coherent impulse responses from the \ac{RU} positions but allows a quantitative assessment of properties such as the uplink energy efficiency \cite{Choi21a}. Furthermore, \ac{UE} mobility and scatterer mobility cannot be captured by the distributed virtual array measurement principle of \cite{Choi21}.

\subsection{Contributions of this work}
\begin{itemize}
	\item World-first fully parallel and coherent widely distributed array channel sounding measurements in a high mobility scenario with 32 \ac{BS} antennas and two users.
	\item Comparative analysis of wireless propagation conditions for \ac{BS} aperture sizes from \SI{2}{\meter} to \SI{46.5}{\meter} in terms of delay spread, Doppler spread, received power, and collinearity of the \ac{LSF}.
	\item Evaluation of channel hardening and effects of channel aging on the \ac{SINR} in a cell-free massive \ac{MIMO} system based on the obtained measurement data.
\end{itemize}

\subsection{Organization of this work}
The paper is organized as follows. \Cref{sec:ScenarioDescriptionAndMeasurementFramework} describes the measurement scenario and the measurement framework utilized for this purpose. In \Cref{sec:ChannelSoundingSignalModel}, the signal model for the obtained data is introduced and the necessary parameters for evaluation are defined. \Cref{sec:ResultsOfStatisticalCharacterization} presents the corresponding results. \Cref{sec:MassiveMIMOProcessingAndResults} introduces the signal model for a cell-free massive \ac{MIMO} system and describes the effects of channel aging and channel hardening, based on the obtained measurement results.


\section{Scenario description and measurement framework }
\label{sec:ScenarioDescriptionAndMeasurementFramework}
We report a widely distributed massive \ac{MIMO} channel sounding campaign that was conducted in March 2022 at the premises of AIT Austrian Institute of Technology GmbH in Vienna, Austria. In this campaign, three different array geometries with apertures ranging from \SI{2}{\meter} to \SI{46.5}{\meter} were installed and tested. We thereby analyze the merits and drawbacks of widely distributed and cell-free systems in comparison to conventional massive \ac{MIMO} systems currently being deployed and operational worldwide.

Massive \ac{MIMO} systems are especially suitable for dealing with urban environments and multipath propagation, since the large number of \acp{RU} allows for spatial focusing and spatial separation of users. However, mobility is still problematic, as it generally leads to outdated \ac{CSI} for beam-forming when not properly accounted for \cite{Loeschenbrand2020}. 

Cell-free systems are envisioned to be deployed in urban environments, as they potentially mitigate the burden of large-scale fading, \ie{,} blocking by buildings, vegetation, cars etc. 
The cell-free massive \ac{MIMO} channel sounding campaign is designed to capture these urban channel characteristics, including mobility, multipath propagation, blocking and transition from  \ac{LOS} to \ac{NLOS}. 

Two vertically polarized monopole transmit antennas are mounted on the rooftop of a van right above the driving seat and the passenger seat, respectively. They are acting as \acp{UE} transmitting a signal and are referenced in the following as $\mathrm{UE}_\indMS$ with $\indMS\in\{1,2\}$. {Users $\mathrm{UE}_\indMS$ are following a fixed trajectory for all measurements, with velocities ranging from \SI{15}{\kilo\meter\per\hour} to \SI{60}{\kilo\meter\per\hour}.} The trajectory is divided into eight regions $\mreg{s},\ s\in\{1,\ \ldots, 8\}$, with a length of approximately \SI{40}{\meter} each, for easy referencing and location-specific data evaluation. \Cref{fig:ScenarioOverview} shows a top view of the scenario under consideration, with the \ac{UE} trajectory and the regions \reg{1} to \reg{8} indicated in blue. The starting position of the \acp{UE} is marked with a white van icon.

\Figure[!h]()[width=\textwidth]{graphics/ScenarioOverview.tikz}{Top view of the measurement scenario. The \ac{UE} trajectory is divided into regions \reg{1} to \reg{8} and indicated in blue. The \ac{BS} antenna array is located on the roof of an office building and indicated in green. The individual \acp{RU} are facing a large office building to the north, dividing the \ac{UE} trajectory into \ac{LOS} and \ac{NLOS} regions. \label{fig:ScenarioOverview}}

On the \ac{BS} side, 32 \acp{RU} consisting of single patch antennas are positioned on the roof top of an office building at a height of \SI{15}{\meter} as a horizontal linear array, with vertical polarization and their individual main lobe facing north. The \acp{RU} are receiving the signal transmitted by the \acp{UE}. The green area in \cref{fig:ScenarioOverview} shows the position of the \ac{BS} array. Since there is an office building of similar height to the north of the linear \ac{BS} antenna array (in the directions the individual patch antenna array elements are pointing to), regions \reg{3} to \reg{5} of the \ac{UE} trajectory are exhibiting \ac{NLOS} conditions. Regions \reg{1} and \reg{2} are characterized as \ac{LOS} with the main direction of \ac{UE} movement perpendicular to the \ac{BS} antenna array, while regions \reg{6}, \reg{7}, and \reg{8} show \ac{LOS} characteristics with the main direction of \ac{UE} movement parallel to the \ac{BS} antenna array. The region where the blocking building causes \ac{NLOS} conditions and the transition to \ac{LOS} is indicated as a gray shade in \cref{fig:ScenarioOverview}.

Three different linear horizontal \ac{BS} array configurations are implemented, with significant variation in the array aperture and antenna element spacing. \Cref{fig:RXconfigs2} shows in green the position where all three array configurations are located. As a reference point, the easternmost antenna 1 stayed on the same position throughout all measurements. The white elements in the green bars at the top of \cref{fig:RXconfigs2} indicate the individual position and spacing of antenna elements for the three array configurations and will serve as reference and visualization aid throughout the remainder of this paper. The exact positioning and spacing of the patch antenna elements is detailed in the next section.


\Figure[!h]()[width=0.1\linewidth]{graphics/RXconfigs2.tikz}{Top view of the roof where three different \ac{BS} array configurations are realized. The white elements in the green bars at the top indicate the individual position and spacing of antenna elements for the three array configurations. \label{fig:RXconfigs2}}

\subsection{Flexible wide aperture array measurement framework}
\label{sec:FlexibleWideApertureArrayMeasurementFramework}
A measurement framework based on our previous work in \cite{Loschenbrand2019} is used to capture the sounding sequences emitted from users $\mathrm{UE}_\indMS$. In addition to each of the two \acp{DU} having a distance of up to \SI{30}{\meter} to the central \ac{BPU} and synchronization unit (as outlined in \cite{Loschenbrand2019}), coaxial cables with a length of \SI{15}{\meter} are used to connect each \ac{DU} with 16 \acp{RU} (patch antenna array elements), see \cref{fig:RXconfigs} at the bottom. With this method, array apertures of up to \SI{90}{\meter} with 32 array elements are possible, all while maintaining fully parallel and fast channel sounding capabilities. To compensate for the losses of the long coaxial cables, power amplifiers with a \SI{1}{\decibel} compression point of \SI{39}{\deci\belm} are used at the UE side.

\Figure[!h]()[width=\textwidth]{graphics/RXconfigs.tikz}{Channel measurement framework consisting of a central \ac{BPU} for data aggregation and storage (bottom), two \acp{DU} for up- and down conversion, and 32 \acp{RU} assembled in one of three \ac{BS} array configurations (top).\label{fig:RXconfigs}}

With the \ac{BS} setup as described above, different antenna array geometries can be realized with minimal effort. We implement three different horizontal linear arrays with varying aperture size and element spacing. Detailed array dimensions are provided in \cref{fig:RXconfigs}.

\textit{\ac{BS} array configuration 1 (\BSconf{} 1)} resembles a conventional linear massive \ac{MIMO} array with 32 antenna elements aligned horizontally and spaced $0.64\lambda$ with $\lambda$ being the wavelength at the sounding frequency of $f=\SI{3.2}{\giga\hertz}$. With this array configuration, grating lobes are mostly avoided and the transmitting \acp{UE} operate in the far-field \cite[Sec.~4]{Molisch2009} of the \ac{BS} array, \ie{,} the distance $d_\indMS$ between $\mathrm{UE}_\indMS$ and \ac{BS} array satisfies
\begin{equation}
	d_\indMS\geq\frac{2D^2}{\lambda}{,}
	\label{eqn:FarField}
\end{equation}
with $D$ being the size of the linear \ac{BS} array. 
\oldnew{Large-scale fading and other channel statistics are expected to be very similar over the \ac{BS} array aperture.}{Therefore, all 32 \ac{BS} antennas operate in an almost identical propagation environment and exhibit strong similarities in path-loss, blocking by large objects, visibility of main scatter sources, and relative \ac{UE} velocity.}

\textit{\ac{BS} array configuration 2 (\BSconf{} 2)} resembles a distributed massive \ac{MIMO} array with two antenna arrays of 16 elements. The two arrays are distributed with a distance of \SI{44.6}{\meter} between them. Each of the distributed arrays is again assembled with horizontally aligned and $0.64\lambda$-spaced antenna elements. With this distributed array configuration and a total aperture size of \SI{46.5}{\meter}, the \acp{UE} never operate in the far-field of the \ac{BS} array according to \cref{eqn:FarField}. This implies that wavefronts impinging at the \ac{BS} array are spherical in general and not of equal amplitude. \oldnew{Large-scale fading and other channel statistics are expected to be very similar for each distributed 16-element array on its own, but may differ greatly from one distributed array to the other.}{The elements of a distributed 16-element array operate in an almost identical propagation environment. But propagation characteristics like path-loss, shadowing, and Doppler due to the relative \ac{UE} velocity can differ substantially from one distributed array to the other.}

\textit{\ac{BS} array configuration 3 (\BSconf{} 3)} resembles a cell-free massive \ac{MIMO} setup with 32 antenna elements horizontally aligned, but spaced with a distance of $16\lambda$ from one array element to the next. {This configuration can be considered as implementation of the radio stripe architecture proposed in \cite{interdonato2019ubiquitous}.} \oldnew{Since the large aperture of \SI{46.5}{\meter} is uniformly filled with receiving elements, each one of them is expected to exhibit variations in large-scale fading and other channel statistics due to their large spatial separation.}{Since receiving elements are uniformly distributed over a large aperture of \SI{46.5}{\meter}, they exhibit strong variations of propagation characteristics such as path-loss, shadowing, and Doppler due to the relative \ac{UE} velocity.}

The parameters of the channel sounding framework utilized for the presented measurement campaign are listed in \Cref{tab:MeasurementParameters}. For each \ac{BS} array configuration, 10 runs on the same \ac{UE} trajectory with similar velocities are performed. One measurement run lasts \SI{30}{\second}, in which a distance of around \SI{300}{\meter} (divided in eight regions) is covered.
\begin{table}[htbp]
	\caption{Measurement parameters and their respective values.}
	\begin{center}
		\begin{tabular}{ll} 
			\toprule
			{Parameter} 				& {Value} \\ 
			\midrule
			carrier frequency $f$ 		& $\SI{3.2}{\giga\hertz}$ \\  
			number of tones $\numTones$ & $481$\\  
			tone spacing $\diffFreq$ 	& $\SI{240}{\kilo\hertz}$ \\  
			bandwidth $B$				& $\SI{115.44}{\mega\hertz}$  \\
			repetition rate $T_\mathrm{R}$ & $\SI{1}{\milli\second}$\\  
			max.\ resolvable Doppler frequency 	& $\SI{500}{\hertz}$\\
			max.\ resolvable velocity 				& $\SI{150}{\kilo\meter\per\hour}$\\
			TX power 					& $\SI{38}{dBm}$\\
			measurement time per run	& $\SI{30}{\second}$\\
			runs per \ac{BS} config.	& $10$\\
			\acp{UE} distance traveled	& $\SI{300}{\meter}$\\
			\bottomrule
		\end{tabular} 
		\label{tab:MeasurementParameters}
	\end{center}
\end{table}

\section{Channel Sounding Signal Model}
\label{sec:ChannelSoundingSignalModel}

With the channel sounding framework described above, we obtain bandlimited estimates of the wireless channel transfer function by transmitting a known sounding sequence from $\numMS=2$ users $\mathrm{UE}_k$, receiving it at the $\numBS=32$ \acp{RU} simultaneously and storing the sampled time signal on a hard-drive. In post-processing, this stored time signals are Fourier-transformed and calibrated to correct for the effects of the \ac{RF} chain. The obtained time-dependent realizations of the channel matrix are then used to derive and analyze the channel characteristics. The procedure is detailed in the following (see also \cite{Loschenbrand2019,Loschenbrand2019a}).

We use a complex baseband multitone signal\oldnew{ }{, defined as}
\begin{equation}
	x[n] = x(nT/Q) = \sum_{\indTones=-(\numTones-1)/2}^{(\numTones-1)/2}X[\indTones] e^{i2\pi\indTones n / \numTones}{,}
	\label{eqn:MultitoneSignal}
\end{equation}
to capture the channel characteristic over time. The time signal $x[n]$ is formed by a superposition of $\numTones$ tones with complex weights $X[\indTones]$ and frequency spacing $\diffFreq$. \oldnew{The amplitudes of the complex weights are chosen to be equal, \ie $|X[\indTones]|=1$.}{The amplitudes of the complex weights $|X[\indTones]|=1$ are chosen to be equal.} Their phases are optimized to achieve a low crest factor \cite{Andreas2011a}
\begin{equation}
	C = \frac{\max_{t\in\left[0,T\right]} |x(t)|}{\sqrt{\frac{1}{N}\int_{0}^{T}|x(t)|^2\diffd t}}
	\label{eqn:CrestFactor}
\end{equation}
of the continuous signal $x(t)$ over the period $T=1/\diffFreq$ with the algorithm proposed in \cite{Friese1997}.

The multitone signal $x[n]$ utilized in this work features a frequency spacing of $\diffFreq=\SI{240}{\kilo\hertz}$ and $\numTones=481$ tones, thus yielding a bandwidth of $B = \SI{115.44}{\mega\hertz}$, with a crest factor $C = 1.24$. Similar to an \ac{OFDM} system, we concatenate the multitone signal with a copy of itself which acts as \ac{CP}. Additionally, a third copy is added to increase the \ac{SNR}. Thus, the final sounding signal consists of three concatenated copies of the multitone signal $x[n]$ and therefore lasts $3\, T = \SI{12.5}{\micro\second}$. Each user $\mathrm{UE}_\indMS$ transmits the same sounding sequence with a $(\indMS-1)3\, T$ time shift to not interfere with the current measurement.

The sounding signal is sent and received with a repetition rate of $T_{\mathrm{R}}=\SI{1}{\milli\second}$, which results in a maximum resolvable Doppler frequency of \SI{500}{\hertz} and a maximum resolvable \ac{UE} velocity of \SI{150}{\kilo\meter\per\hour} at a sounding frequency of $f=\SI{3.2}{\giga\hertz}$.

Each individual antenna $\indBS \in \{1, 2,\ \ldots, \numBS\}$ of the \ac{BS} receives the sounding signal (convolved with the propagation channel $\hat{H}_{\indBS,\indMS}[m,q] $) $\tilde{y}_{\indBS,\indMS}[m,n]$, transmitted by $\mathrm{UE}_\indMS$. Since the sounding procedure works similar to a \ac{CP} \ac{OFDM} scheme, the received multitone signal weights are obtained by omitting the \ac{CP} and guard periods in $\tilde{y}_{\indBS,\indMS}[m,n]$ to obtain ${y}_{\indBS,\indMS}[m,n]$ and a subsequent discrete Fourier transformation, \ie{,}
\begin{equation}
	Y_{\indBS,\indMS}[m,\tilde{\indTones}] = \frac{1}{\sqrt{2\numTones}}\sum_{n=0}^{2\numTones-1} {y}_{\indBS,\indMS}[m,n]\, \exp^{-\frac{j2\pi n\tilde{\indTones}}{2\numTones}},
\end{equation}
where $\indDelay$ denotes discrete time and $\indTones$ discrete frequency.

\oldnew{The signal part of the sounding sequence contains two repetitions of the multitone signal \cref{eqn:MultitoneSignal}. Therefore, every second bin $Y_{\indBS,\indMS}[m,\indTones] = Y_{\indBS,\indMS}[m,2\tilde{\indTones}]$ constitutes the uncalibrated transfer function of the propagation channel and the radio front-ends. The transmitted symbols $X[\indTones]$ are known at the receiver and the calibrated transfer function estimate of the channel is thus obtained by
\begin{equation}
	\hat{H}_{\indBS,\indMS}[m,q] = \frac{Y_{\indBS,\indMS}[m,\indTones]}{X[\indTones]\hat{H}_{\indBS,\indMS}^{\mathrm{RF}}[q]}.
	\label{eqn:TFestold}
\end{equation}
The transfer function $\hat{H}_{\indBS,\indMS}^{\mathrm{RF}}[q]$ of the \ac{RF} chains is obtained during a calibration phase prior to the measurement \cite{Loschenbrand2019}.}
{The signal part of the sounding sequence contains two repetitions of the multitone signal \cref{eqn:MultitoneSignal}. Therefore, every second frequency bin $Y_{\indBS,\indMS}[m,\indTones] = Y_{\indBS,\indMS}[m,2\tilde{\indTones}]$ constitutes the uncalibrated transfer function of the propagation channel and the radio front-ends. We consider the transmitted symbols $X[\indTones]$ in frequency domain as pilot symbols which are known at the receiver. Thus, we calculate the calibrated transfer function estimate of the channel by a \ac{LS} estimation \cite[Sec.~1.7]{Coleri2002,Hlawatsch11} as
\begin{equation}
	\hat{H}_{\indBS,\indMS}[m,q] = \frac{Y_{\indBS,\indMS}[m,\indTones]}{X[\indTones]\hat{H}_{\indBS,\indMS}^{\mathrm{RF}}[q]}.
	\label{eqn:TFest}
\end{equation}
The transfer function $\hat{H}_{\indBS,\indMS}^{\mathrm{RF}}[q]$, representing the transmit and receive \ac{RF} chain properties, is obtained during a calibration phase prior to the measurement \cite{Loschenbrand2019}.}

\subsection{Statistical characterization of the wireless propagation channel}
\label{subsec:StatisticalCharacterizationOfTheWirelessPropagationChannel}

To draw conclusions about wide-sense stationarity or the lack thereof for the \ac{BS} array configurations under investigation, we revert to analyzing the moments of the wireless channels' {time-variant} \ac{PDP} and \ac{DSD}. To that end, the \ac{LSF} is introduced as it provides a means of calculating the considered parameters in highly dynamic scenarios \cite{Matz2005,Paier08}. {The \ac{LSF} characterizes the dispersion in Doppler and delay.}
%

A certain number of consecutive measurements $\numDelay$ {within a local stationarity region} is necessary to calculate the \ac{LSF} in \cref{eq:LSF}. Therefore, \oldnew{a}{}discrete time $l$ is introduced to index the calculated \acp{LSF} and its moments and marginals. The time index $l$ is often referred to in literature as the stationarity region index, while $\numDelay$ is then the stationarity region length \cite{Matz2005,Bernado14} and chosen to be even in this work.

Given the time-variant frequency transfer function estimate $\hat{H}_{\indBS,\indMS}[\indDelay ,\indTones ]$ from $\mathrm{UE}_\indMS$ to \ac{BS} antenna $\indBS$, the estimate of the \ac{LSF} at stationarity region index $l$ is given as {\cite{Matz2005,Percival93}}
\begin{equation}
	\hat{\mathcal{C}}_{\indBS,\indMS}[l;n,p]=\frac{1}{IJ}\sum_{w=0}^{IJ-1}\left\vert\mathcal{H}_{\indBS,\indMS}^{(G_w)}[l;n,p]\right\vert^2{,}
	\label{eq:LSF}
\end{equation}
with the Doppler shift index $p\in \{-\numDelay/2,\ \ldots,\numDelay/2-1\}$ and the delay index $n\in\{0,\ \ldots,\numTones -1\}$. The delay and Doppler shift resolution are defined by $\tau_\text{s}=\frac{1}{\numTones \Delta f}$ and $\nu_\text{s}=\frac{1}{\numDelay T_\text{R}}$, where $T_{\mathrm{R}}$ is the repetition rate of measurement. The operation $\indDelay=l\numDelay+\indDelay'$ maps the measurement time index $\indDelay$ to the stationarity region index $l$. The tapered frequency response is
\begin{align}
	\mathcal{H}_{\indBS,\indMS}^{(G_w)}[l;n,p]&=\sum_{\indDelay =-\numDelay/2}^{\numDelay/2-1}\sum_{\indTones =-(\numTones -1)/2}^{(\numTones -1)/2} \hat{H}_{\indBS,\indMS}[\indDelay' +\numDelay l,\indTones ]\nonumber\\
	&\cdot G_w[\indDelay' ,\indTones ]\mathrm{e}^{-\mathrm{j}2\pi(p\indDelay' -n\indTones )},
\end{align}
where the tapers $G_w[\indDelay ,\indTones ]$ are two-dimensional \ac{DPS} sequences as shown in detail in \cite{Bernado14,Slepian78}. The number of tapers in the time and frequency domain is set to $I=2$ and $J=1$, respectively \cite{Bernado14,Zoechmann2019}.

We calculate the \ac{PDP} and \ac{DSD} as projections of the \ac{LSF} onto the Doppler domain and the delay domain, respectively {\cite{Hlawatsch11,Molisch2009}}:
\begin{align}
	\hat{\mathcal{P}}^{(\tau)}_{\indBS,\indMS}[l;n] &= 
	\frac{1}{\numDelay}\sum_{p=-\numDelay/2}^{\numDelay/2-1}\hat{\mathcal{C}}_{\indBS,\indMS}[l;n,p],
	\label{eq:PDP}\\
	\hat{\mathcal{P}}^{(\nu)}_{\indBS,\indMS}[l;p] &= 
	\frac{1}{\numTones }\sum_{n=0}^{\numTones -1}\hat{\mathcal{C}}_{\indBS,\indMS}[l;n,p].
	\label{eq:DSD}
\end{align}

\subsection{Moments of the \ac{PDP} and \ac{DSD}}
{To allow for a comprehensive  and quantitative analysis of the time-variant \ac{PDP} and \ac{DSD}, we revert to their normalized moments as characterizing parameters \cite[Sec.~6.5]{Molisch2009}, \cite[Sec.~1.4.1.5]{Hlawatsch11}. The following formulas for the normalized moments of \ac{PDP} and \ac{DSD} are taken from \cite{Hlawatsch11} and adopted to the used sampling in time and frequency.}

The time-variant first moment of the \ac{PDP} of the estimated channel transfer function $\hat{H}_{\indBS,\indMS}[m,\indTones ]$, \ie{,} its time-integrated power, is calculated as 
\begin{equation}
	\hat{\mathcal{P}}^{(\tau)}_{\indBS,\indMS}[l] = \sum_{n=0}^{\numTones-1}\hat{\mathcal{P}}_{\indBS,\indMS}^{(\tau)}[l;n].
\end{equation}
The second moment is referred to as the mean delay and calculated as 
\begin{equation}
	\bar{\tau}_{\indBS,\indMS}[l] = \frac{\sum_{n=0}^{\numTones-1}(n\tau_\text{s})\hat{\mathcal{P}}_{\indBS,\indMS}^{(\tau)}[l;n] } {\hat{\mathcal{P}}^{(\tau)}_{\indBS,\indMS}[l]}.
	\label{meandelay}
\end{equation}
We calculate the \ac{RMS} delay spread by
\begin{equation}
	\sigma_{\indBS,\indMS}^{(\tau)}[l] = \sqrt{\frac{\sum_{n=0}^{\numTones-1}(n\tau_\text{s})^2\hat{\mathcal{P}}_{\indBS,\indMS}^{(\tau)}[l;n] } {\hat{\mathcal{P}}^{(\tau)}_{\indBS,\indMS}[l]}-\bar{\tau}[l]^2}.
	\label{delayspread}
\end{equation}

Similarly, with the integrated power
\begin{equation}
	\hat{\mathcal{P}}^{(\nu)}_{\indBS,\indMS}[l] = \sum_{p=-\numDelay/2}^{\numDelay/2-1}\hat{{P}}_{\indBS,\indMS}^{(\nu)}[l;p]{,}
\end{equation}
and the mean Doppler
\begin{equation}
	\bar{\nu}_{\indBS,\indMS}[l] = \frac{\sum_{p=-\numDelay/2}^{\numDelay/2-1}(p\nu_\textbf{s})\hat{\mathcal{P}}_{\indBS,\indMS}^{(\nu)}[l;p]} {\hat{\mathcal{P}}^{(\nu)}_{\indBS,\indMS}[l]}{,}
	\label{meanDoppler}
\end{equation}
we obtain the \ac{RMS} Doppler spread by
\begin{equation}
	\sigma_{\indBS,\indMS}^{(\nu)}[l] = \sqrt{\frac{\sum_{p=-\numDelay/2}^{\numDelay/2-1}(p\nu_\text{s})^2\hat{\mathcal{P}}_{\indBS,\indMS}^{(\nu)}[l;p]} {\hat{\mathcal{P}}^{(\nu)}_{\indBS,\indMS}[l]}-\bar{\nu}_{\indBS,\indMS}[l]^2}.
	\label{Dopplerspread}
\end{equation}

We define the time-dependent average received power at time instant $l$ as the sum over the \ac{LSF} in both time and frequency \cite{Matz2005}
\begin{align}
	\hat{\mathcal{P}}_{\indBS,\indMS}[l] &= \frac{1}{\numDelay \numTones} \sum_{p=-\numDelay/2}^{\numDelay/2-1} \sum_{n=0}^{\numTones -1}\hat{\mathcal{C}}_{\indBS,\indMS}[l;n,p]\\
	&= \frac{1}{\numTones} \hat{\mathcal{P}}^{(\tau)}_{\indBS,\indMS}[l]= \frac{1}{\numDelay} \hat{\mathcal{P}}^{(\nu)}_{\indBS,\indMS}[l].
\end{align}
Let $L(s)$ denote the set of time indices $l$ for which the user $\mathrm{UE}_\indMS$ moves in a given region $\mreg{s},\ s\in\{1,\ \ldots, 8\}$ (see also \cref{fig:ScenarioOverview}). If user $\mathrm{UE}_\indMS$ is in region $\mreg{s}$ at time instant $l$, then $l\in L(s)$. The size of the set $L(s)$, representing the time a user spent in a given region, is denoted by $|L(s)|$. The average received power from user $\mathrm{UE}_\indMS$ in region $\mreg{s}$ is then defined as
\begin{equation}
	\hat{\mathcal{P}}_{\indBS,\indMS}[s] = \frac{1}{|L(s)|}\sum_{l\in L(s)} \hat{\mathcal{P}}_{\indBS,\indMS}[l].
\end{equation}
For a more convenient exposition later on, we also define the normalized received power from user $\mathrm{UE}_\indMS$ in region $\mreg{s}$ by dividing with the maximum value of $\frac{1}{\numBS}\sum_{\indBS=1}^{\numBS}\hat{\mathcal{P}}_{\indBS,\indMS}[s]$ over all regions $\mreg{s},\ s\in\{1,\ \ldots, 8\}$,
\begin{equation}
	\bar{\mathcal{P}}_{\indBS,\indMS}[s] = \frac{\hat{\mathcal{P}}_{\indBS,\indMS}[s]}{\max_{s}\frac{1}{\numBS}\sum_{\indBS=1}^{\numBS}\hat{\mathcal{P}}_{\indBS,\indMS}[s]}. \label{eqn:normAvrgRXpower}
\end{equation}

\subsection{Collinearity of the \ac{LSF}}
We define the collinearity \cite{Paier08,Bernado14} of the \ac{LSF} at time $l$ as 
\begin{equation}
	\gamma_{\indBS,\indBS',\indMS}[l] = \frac{\sum_{p=-\numDelay/2}^{\numDelay/2-1} \sum_{n=0}^{\numTones -1} \hat{\mathcal{C}}_{\indBS,\indMS}[l;n,p] \odot \hat{\mathcal{C}}_{\indBS',\indMS}[l;n,p]}
	{\left\lVert\hat{\mathcal{C}}_{\indBS,\indMS}[l;n,p]\right\rVert_F \left\lVert\hat{\mathcal{C}}_{\indBS',\indMS}[l;n,p]\right\rVert_F},
	\label{eqn:LFSCollinearity}
\end{equation}
with 
\begin{equation}
	\left\lVert\hat{\mathcal{C}}_{\indBS,\indMS}[l;n,p]\right\rVert_F = \sqrt{\sum_{p=-\numDelay/2}^{\numDelay/2-1} \sum_{n=0}^{\numTones -1} \left|\hat{\mathcal{C}}_{\indBS,\indMS}[l;n,p]\right|^2}
\end{equation} denoting the Frobenius Norm of the \ac{LSF}, to evaluate the stationarity of the three proposed \BSconf{} 1 to 3 in space. We thereby analyze, if the wireless propagation channel to different \acp{RU} exhibits different statistical parameters. A value close to one indicates a similar distribution of multipath components in the delay-Doppler domain at a certain time (or equivalently in space). In contrast, a value close to zero indicates no similarity in the multipath component distribution.

The average spatial collinearity between \acp{RU} over $l$ is defined as
\begin{equation}
	\bar{\gamma}_{\indBS,\indBS',\indMS} = \frac{1}{\sum_{s=1}^{8}|L(s)|}\sum_{l}\gamma_{\indBS,\indBS',\indMS}[l]
	\label{eqn:avrgLFSCollinearity}
\end{equation}
and is used to assess the average value of \ac{LSF} collinearity over the full \ac{UE} trajectory. 
\section{Results of statistical characterization}
\label{sec:ResultsOfStatisticalCharacterization}
In this section, we analyze the data gathered as shown in \cref{sec:FlexibleWideApertureArrayMeasurementFramework} with the methods outlined in \cref{subsec:StatisticalCharacterizationOfTheWirelessPropagationChannel}. We seek knowledge of the characteristics of each channel from user $\mathrm{UE}_1$ to \ac{BS} antenna $\indBS$, and their relation and correlation to each other. Although ten measurement runs were obtained for each \ac{BS} array configuration, the results of only one run per configuration are presented for the ease of exposition. Including more measurement runs has been analyzed and did not show significant differences in the presented results.
\renewcommand{\indMS}{1}

\subsection{Average received power}
We evaluate the average received power on each individual \ac{BS} antenna from transmitter $\mathrm{UE}_1$ for the three considered \ac{BS} array configurations along the \ac{UE} trajectory. Results are normalized to the maximum average received power as defined in \cref{eqn:normAvrgRXpower}, which occurred in region \reg{6} with \BSconf{} 1. The height of the 32 bars per region indicates $10\log_{10}\bar{\mathcal{P}}_{\indBS,\indMS}[s]$ and the color indicates the distance to \ac{RU} 1 for each antenna in the \ac{BS} array. The average normalized received power over all \acp{RU} within a region $10\log_{10}\frac{1}{\numBS}\sum_{\indBS=1}^{\numBS}\hat{\mathcal{P}}_{\indBS,\indMS}[s]$ is provided at the base of the bar plots for each region.

\Cref{fig:receivedPower_SC1_R2_reg1-8_A1-32} shows the average normalized received power $\bar{\mathcal{P}}_{\indBS,\indMS}[s]$ for regions \reg{1} to \reg{8} and \BSconf{} 1. Since this configuration only has a small overall aperture size of approximately \SI{2}{\meter} (see also \cref{fig:RXconfigs}), all 32 \acp{RU} experience the same large-scale fading and $\bar{\mathcal{P}}_{\indBS,\indMS}[s]$ is similar within each region. In the \ac{NLOS} regions \reg{3} to \reg{5}, the average received power drops \SI{15}{} - \SI{20}{\decibel} compared to the strong \ac{LOS} case in region \reg{6}.

\Figure[!h]()[width=0.98\columnwidth]{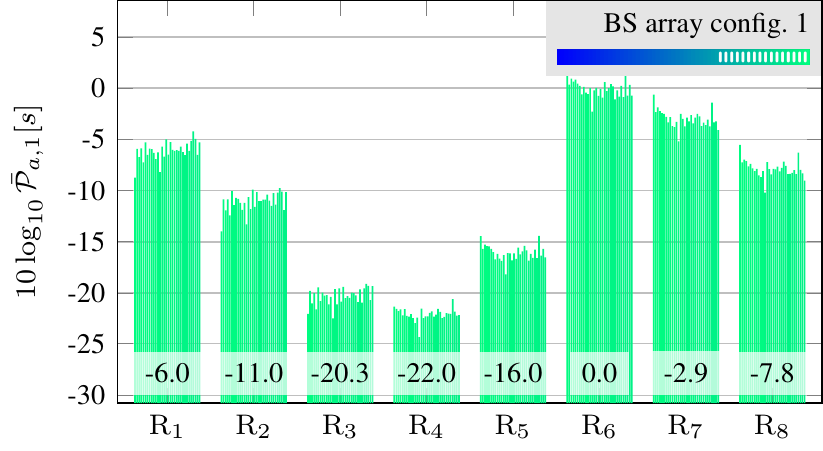}{Average received power $\bar{\mathcal{P}}_{\indBS,\indMS}[s]$ for \BSconf{} 1 and regions \reg{1} to \reg{8}. Each regions contains 32 bars for all \acp{RU}. The color coding shows the respective distance to \ac{BS} antenna 1 as indicated in the top right corner. \label{fig:receivedPower_SC1_R2_reg1-8_A1-32}}

In contrast to the previous result, \cref{fig:receivedPower_SC2_R2_reg1-8_A1-32} with \BSconf{} 2 shows significant differences in large-scale fading along the aperture of \SI{46.5}{\meter}. It is obvious that in this case, the common assumption of wide-sense stationarity among \acp{RU} is no longer given. The difference in average received power between the \ac{BS} antenna groups (see also \cref{fig:RXconfigs}) is typically \SI{7}{} - \SI{8}{\decibel} (essentially the free space path-loss between \ac{BS} antenna groups), but occasionally grows as large as \SI{20}{\decibel} like in region \reg{6}. In this case, \acp{RU} 1-16 have direct \ac{LOS}, whereas antennas 17-32 are still blocked by the large office building to the north of the \ac{BS} array. Only in region \reg{4} with \ac{NLOS} propagation conditions and approximately the same distance from \ac{UE} node to all \acp{RU} the average received power stays on a similar level over the \ac{BS} aperture.

\Figure[!h]()[width=0.98\columnwidth]{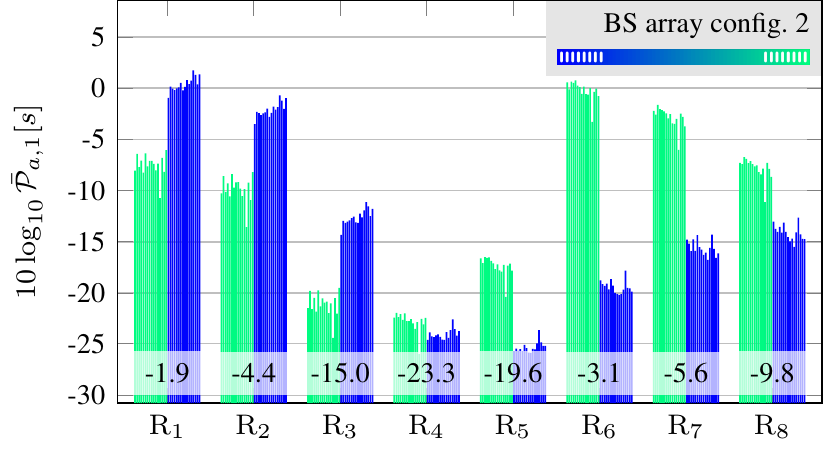}{Average received power $\bar{\mathcal{P}}_{\indBS,\indMS}[s]$ for \BSconf{} 2 and regions \reg{1} to \reg{8}. Each regions contains 32 bars for all \acp{RU}. The color coding shows the respective distance to \ac{BS} antenna 1 as indicated in the top right corner. \label{fig:receivedPower_SC2_R2_reg1-8_A1-32}}

Similar to before, \BSconf{} 3 shows strong non-stationarity over its aperture of \SI{46.5}{\meter} as shown in \cref{fig:receivedPower_SC3_R2_reg1-8_A1-32}. Again, the differences in average received power of \SI{7}{} - \SI{8}{\decibel} in regions \reg{1} to \reg{3} and \reg{5} is mostly explained with the different distance from the \ac{UE} to the individual \ac{BS} antenna and the corresponding path-loss. Region \reg{6} shows the average received power with a variation of \SI{20}{\decibel} among \acp{RU} due to a large building blocking parts of the array aperture. 

\Figure[!h]()[width=0.98\columnwidth]{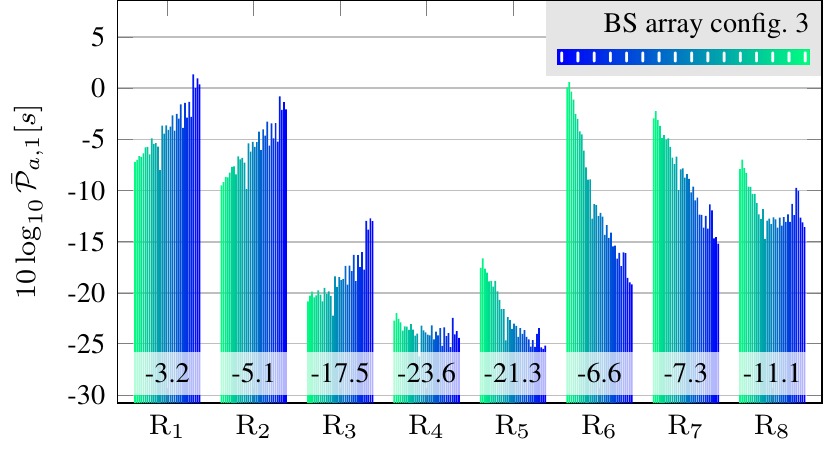}{Average received power $\bar{\mathcal{P}}_{\indBS,\indMS}[s]$ for \BSconf{} 3 and regions \reg{1} - \reg{8}. Each regions contains 32 bars for all \acp{RU}. The color coding shows the respective distance to \ac{BS} antenna 1 as indicated in the top right corner. \label{fig:receivedPower_SC3_R2_reg1-8_A1-32}}

\subsection{Delay- and Doppler spreads}
The large array aperture of \BSconf{} 2 and 3 (in relation to configuration 1) not only causes great variation in the average received power as seen in the previous section. Due to the \acp{UE} operating in the near-field, the relative distances and velocities between \acp{UE} and \acp{RU} vary greatly. To quantify this effect, we revert to the delay spread and Doppler spread as defined in \cref{delayspread} and \cref{Dopplerspread}, respectively. 

A stationarity time of \SI{256}{\milli\second} for the \ac{LSF} was assumed for the necessary calculations, \ie{,} $\numDelay=256$. Comparative studies with a smaller stationarity time (\ie{,} \SI{128}{\milli\second} or \SI{64}{\milli\second}) did not show significant differences in the presented results, but offer less resolution in the Doppler domain (see \cref{eq:LSF}). 

Additionally, thresholds are applied to the delay and Doppler components of the \ac{LSF} to reduce influence of measurement noise and limited dynamic range on the calculation of the spreads. The noise threshold is set to \SI{3}{\decibel} and the sensitivity threshold is set to \SI{45}{\decibel} \cite{Bernado14}.

\paragraph*{Delay Spread - \ac{BS} Array Configuration 1} \Cref{fig:DelaySpreadColor_SC1_R2_A1-32} shows the delay spreads $\sigma_{\indBS,\indMS}^{(\tau)}[l]$ as defined in \cref{delayspread} of the $\numBS=32$ time-dependent channel transfer functions  from $\mathrm{UE}_1$ to the \acp{RU} in configuration 1. The time dependency is translated into space by plotting against the cumulative distance $\mathrm{UE}_1$ traveled at each respective time instance.
The color of each line indicates the distance of the respective antenna element to \ac{BS} antenna 1 as indicated in the top right corner of the plot. The gray area between regions \reg{2} and \reg{6} indicates \ac{NLOS} propagation conditions.

We notice that especially in the \ac{LOS} regions \reg{1}, \reg{2}, and \reg{6} - \reg{8}, the delay spread is very similar among \acp{RU}, which is not surprising given the proximity of the individual array elements. In the \ac{NLOS} regions \reg{3} to \reg{5} it naturally rises in general due to the lack of a strong multipath component. 

\Figure[!h]()[width=0.98\columnwidth]{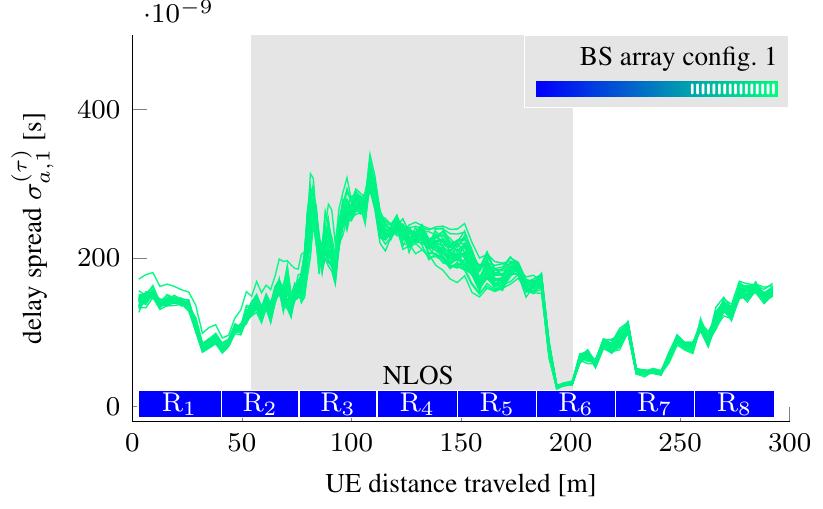}{Delay spread $\sigma_{\indBS,\indMS}^{(\tau)}[l]$ of the $\numBS=32$ time-dependent channel transfer functions  from $\mathrm{UE}_1$ to the \ac{BS} in configuration 1.\label{fig:DelaySpreadColor_SC1_R2_A1-32}}

\paragraph*{Delay Spread - \ac{BS} Array Configuration 2}
\Cref{fig:DelaySpreadColor_SC2_R2_A1-32} shows the delay spreads of the time-dependent channel transfer functions from $\mathrm{UE}_1$ to the \acp{RU} in configuration 2. The edges of the \ac{NLOS} region are indicated in light gray as only half of the \acp{RU} are blocked by the large office building. What is immediately obvious is the clustering of the delay spreads for \acp{RU} in the distributed arrays to the east and west, respectively. In regions \reg{1} to \reg{3} \acp{RU} 1-16 in the east are further from the user $\mathrm{UE}_1$ than antennas 17-32, resulting in an increased delay spread. Regions \reg{6} to \reg{8} show this effect even more pronounced as the \ac{LOS} to \acp{RU} 17-32 is still blocked, resulting in increased delay spreads, whereas antennas 1-16 are close to the UE trajectory and have \ac{LOS}.

\Figure[!h]()[width=0.98\columnwidth]{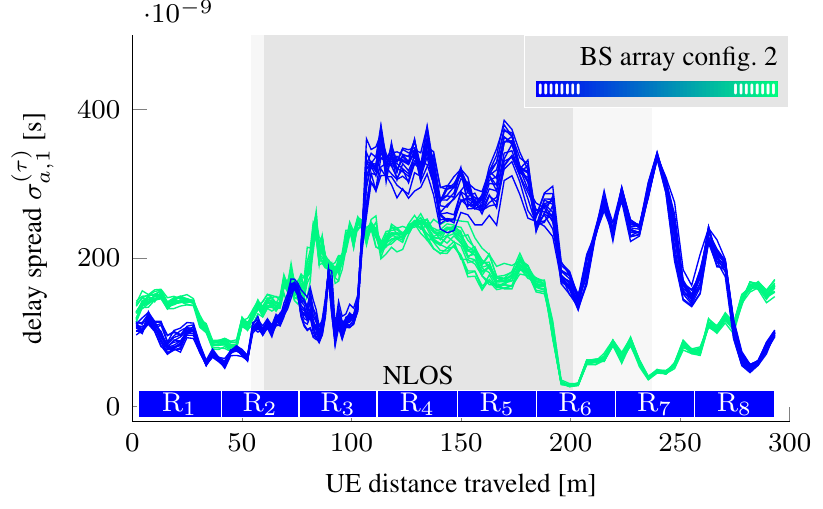}{Delay spread $\sigma_{\indBS,\indMS}^{(\tau)}[l]$ of the $\numBS=32$ time-dependent channel transfer functions  from $\mathrm{UE}_1$ to the \ac{BS} in configuration 2.\label{fig:DelaySpreadColor_SC2_R2_A1-32}}

\paragraph*{Delay Spread - \ac{BS} Array Configuration 3}
\Cref{fig:DelaySpreadColor_SC3_R2_A1-32} shows the delay spreads of the time-dependent channel transfer functions from $\mathrm{UE}_1$ to the \acp{RU} in configuration 3. The edges of the \ac{NLOS} region are fading out as the \ac{LOS} to \ac{NLOS} transition happens gradually over \acp{RU}. Similar to \BSconf{} 2, the range of occurring delay spreads for a given UE position varies widely compared to \BSconf{} 1. But given the uniform distribution of antenna elements, no clustering of delay spreads as in \cref{fig:DelaySpreadColor_SC2_R2_A1-32} is observed. Special emphasis is put on region \reg{6}, where we observe a sudden drop of the delay spread of \SI{150}{\nano\second} consecutively for the \acp{RU}. This effect is caused by $\mathrm{UE}_1$ emerging from the blocking building and \ac{LOS} propagation conditions occurring for one \ac{BS} antenna after the other.

\Figure[!h]()[width=0.98\columnwidth]{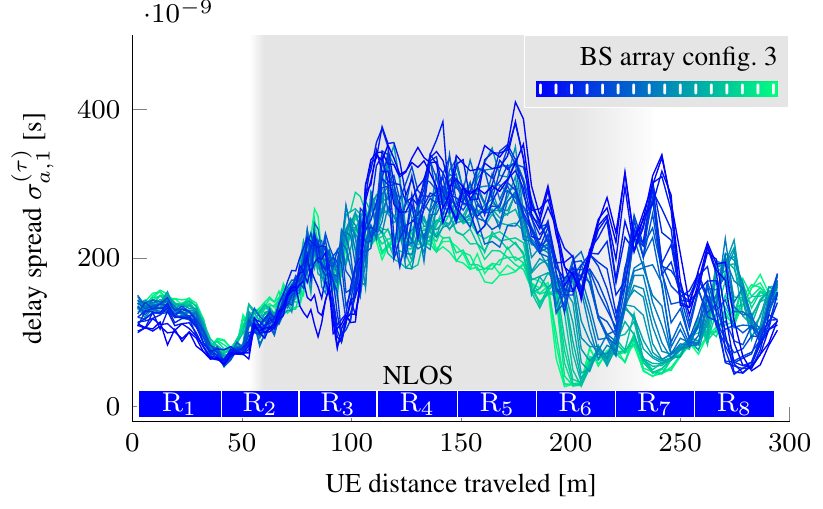}{Delay spread $\sigma_{\indBS,\indMS}^{(\tau)}[l]$ of the $\numBS=32$ time-dependent channel transfer functions  from $\mathrm{UE}_1$ to the \ac{BS} in configuration 3.\label{fig:DelaySpreadColor_SC3_R2_A1-32}}

\paragraph*{Delay Spread - Summary}
\Cref{tab:DelaySpreadStats} shows the quantitative analysis of the qualitative plots presented in \cref{fig:DelaySpreadColor_SC1_R2_A1-32,fig:DelaySpreadColor_SC2_R2_A1-32,fig:DelaySpreadColor_SC3_R2_A1-32}. For each region \reg{1} to \reg{8} and each \BSconf{} 1 to 3, the maximum, minimum, mean, and standard deviation value of the delay spread over the 32 \ac{BS} antenna realizations is provided.

We observe that the maximum, minimum, and mean delay spread values differ at most \SI{20}{\nano\second} for \BSconf{} 2 and 3, as their array aperture and position are identical (\cref{fig:RXconfigs2}). The standard deviation, however, is consistently higher for the distributed massive \ac{MIMO} setup in \BSconf{} 2. The standard deviation and therefore fluctuations in delay spread values among \acp{RU} for \BSconf{} 1 lies between \SI{3}{\nano\second} (\ac{LOS}) and \SI{14}{\nano\second} (\ac{NLOS}) and is in general significantly lower than the respective values for the \BSconf{} 2 and 3, given their significantly larger antenna apertures. 

\begin{table*}\centering
	\caption{Maximum, minimum and mean value of the \ac{RMS} delay spread as well as its standard deviation around the mean within regions \reg{1} to \reg{8} for \BSconf{} 1 to 3.\label{tab:DelaySpreadStats}}
	\begin{tabular}{@{}rrrrrrrrr@{}}\toprule
		& \multicolumn{8}{c}{regions}\\
		\cmidrule{2-9}
		Delay spread $\sigma_{\indBS,\indMS}^{(\tau)}[l]$ & \reg{1} & \reg{2} & \reg{3} & \reg{4} & \reg{5} & \reg{6} & \reg{7} & \reg{8}\\ \midrule
		\ac{BS} array config. 1\\
		Max. [\SI{}{\nano\second}]\ & 181.8& 213.4& 344.9& 317.2& 251.4& 179.5& 133.9& 193.4\\
		Min. [\SI{}{\nano\second}]\ &  72.2&  72.2& 120.2& 164.6&  93.8&  22.8&  37.2&  78.8\\
		Mean [\SI{}{\nano\second}]\ & 119.0& 133.1& 230.2& 229.6& 178.2&  72.6&  73.5& 138.7\\
		Std. [\SI{}{\nano\second}]\ &   5.1&   9.1&  14.0&  11.3&  11.0&   5.3&   3.6&   6.1\\
		\ac{BS} array config. 2\\
		Max. [\SI{}{\nano\second}]\ & 161.9& 194.3& 380.3& 381.6& 388.1& 318.5& 342.2& 245.3\\
		Min. [\SI{}{\nano\second}]\ &  51.7&  51.7&  77.9& 174.8& 147.6&  25.8&  35.4&  45.1\\
		Mean [\SI{}{\nano\second}]\ & 106.4& 119.7& 183.9& 271.9& 239.7& 148.3& 153.2& 129.5\\
		Std. [\SI{}{\nano\second}]\ &  21.3&  14.0&  43.1&  45.7&  60.9&  79.8&  96.0&  42.2\\
		\ac{BS} array config. 3\\
		Max. [\SI{}{\nano\second}]\ & 159.1& 199.9& 383.6& 390.8& 410.8& 315.7& 342.1& 231.0\\
		Min. [\SI{}{\nano\second}]\ &  54.0&  53.4&  76.8& 177.3& 132.4&  25.8&  37.5&  38.0\\
		Mean [\SI{}{\nano\second}]\ & 113.4& 114.9& 199.3& 264.2& 253.9& 141.0& 134.3& 130.8\\
		Std. [\SI{}{\nano\second}]\ &  11.7&  11.7&  33.7&  39.2&  44.8&  57.2&  67.5&  36.2\\
		\bottomrule
	\end{tabular}
\end{table*}

\paragraph*{Doppler Spread - \ac{BS} Array Configuration 1}
\Cref{fig:DelaySpreadColor_SC3_R2_A1-32} shows the Doppler spread of the time-dependent channel transfer functions from $\mathrm{UE}_1$ to the \acp{RU} in configuration 1. Similar to the delay spread analysis, the time dependency is translated into space by plotting against the cumulative distance $\mathrm{UE}_1$ traveled at each respective time instance.
The color of each line indicates the distance of the respective antenna element to \ac{BS} antenna 1 as indicated in the top right corner of the plot.

As with the delay spread, also the Doppler spread variation among \acp{RU} is small at maximally \SI{10}{\hertz} in the \ac{LOS} regions \reg{1}, \reg{2}, and \reg{6} to \reg{8}. The Doppler spread generally increases in the \ac{NLOS} regions \reg{3} to \reg{5} due to the lack of a strong \ac{LOS} component. The relative velocity from $\mathrm{UE}_1$ to the \ac{BS} was low since the UE trajectory hardly showed radial components in these regions, see \cref{fig:ScenarioOverview}. However, Doppler spreads increase significantly in regions \reg{7} and \reg{8} since here the radial velocity component of the UE in relation to the \ac{BS} increases.

\Figure[!h]()[width=0.98\columnwidth]{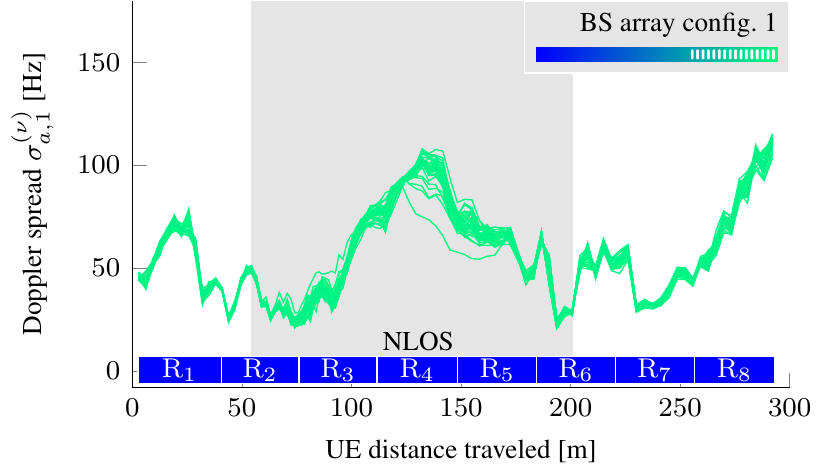}{Doppler spread $\sigma_{\indBS,\indMS}^{(\nu)}[l]$ of the $\numBS=32$ time-dependent channel transfer functions  from $\mathrm{UE}_1$ to the \ac{BS} in configuration 1.\label{fig:DopplerSpreadColor_SC1_R2_A1-32}}

\paragraph*{Doppler Spread - \ac{BS} Array Configuration 2}
\Cref{fig:DelaySpreadColor_SC3_R2_A1-32} shows the Doppler spread of the time-dependent channel transfer functions from $\mathrm{UE}_1$ to the \acp{RU} in configuration 2. Clear clustering can be observed as \acp{RU} 1-16 experience propagation and large-scale fading conditions that are different from \acp{RU} 17-32. Within each distributed group, the Doppler spread is similar among \acp{RU}. Its value is largely determined by the absence of a \ac{LOS} component for the respective antenna group (regions \reg{3} to \reg{5}) and the relative velocity of the UE in respect to the antenna group (regions \reg{6} to \reg{8}).

\Figure[!h]()[width=0.98\columnwidth]{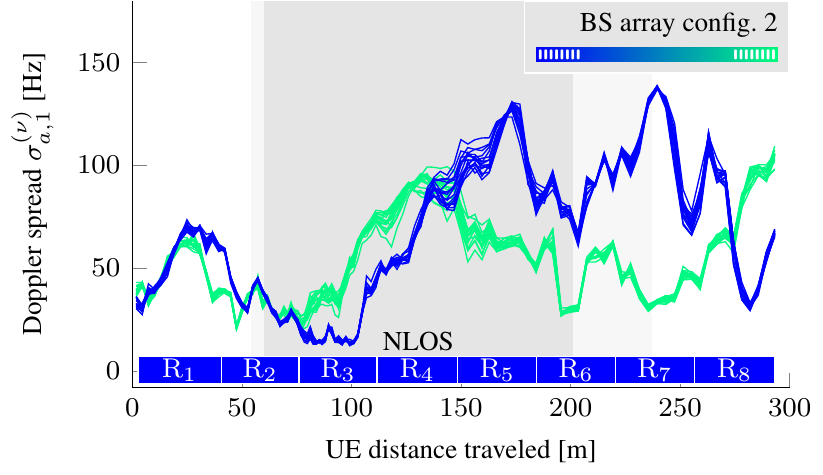}{Doppler spread $\sigma_{\indBS,\indMS}^{(\nu)}[l]$ of the $\numBS=32$ time-dependent channel transfer functions  from $\mathrm{UE}_1$ to the \ac{BS} in configuration 2.\label{fig:DopplerSpreadColor_SC2_R2_A1-32}}

\paragraph*{Doppler Spread - \ac{BS} Array Configuration 3}
\Cref{fig:DelaySpreadColor_SC3_R2_A1-32} shows the Doppler spread of the time-dependent channel transfer functions from $\mathrm{UE}_1$ to the \acp{RU} in configuration 3. The minimum and maximum Doppler spread values along the trajectory of the user are, except for some outliers in region \reg{8}, similar. However, no clustering is observed. At each position in space along the UE trajectory, the Doppler spread values of the 32 \acp{RU} are distributed uniformly between their minimum and maximum values. Again, sudden drops (or increases) in the Doppler spread are observed consecutively over \acp{RU} (regions \reg{6}, \reg{7}) as strong signal components appear (or disappear) for the respective antenna. 

\Figure[!h]()[width=0.98\columnwidth]{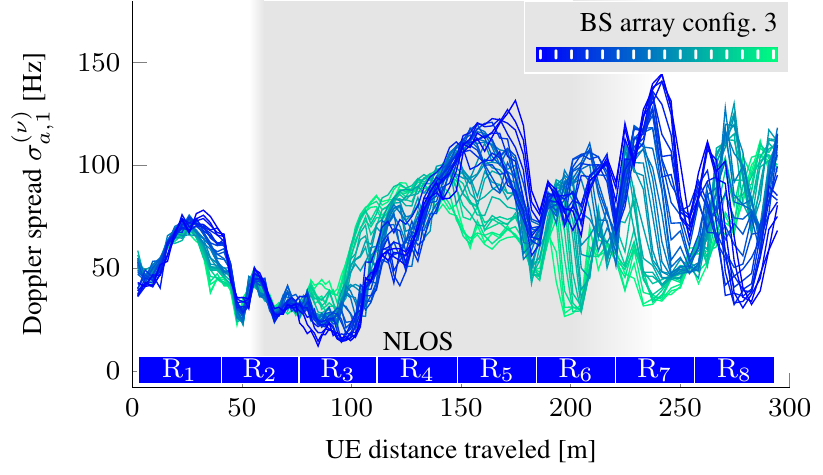}{Doppler spread $\sigma_{\indBS,\indMS}^{(\nu)}[l]$ of the $\numBS=32$ time-dependent channel transfer functions  from $\mathrm{UE}_1$ to the \ac{BS} in configuration 3.\label{fig:DopplerSpreadColor_SC3_R2_A1-32}}

\paragraph*{Doppler Spread - Summary}
\Cref{tab:DopplerSpreadStats} shows the quantitative analysis of the qualitative plots presented in \cref{fig:DopplerSpreadColor_SC1_R2_A1-32,fig:DopplerSpreadColor_SC2_R2_A1-32,fig:DopplerSpreadColor_SC3_R2_A1-32}. For each region \reg{1} to \reg{8} and each \BSconf{} 1 to 3, the maximum, minimum, mean, and standard deviation value of the Doppler spread over the 32 \ac{BS} antenna realizations is provided.

Many findings from analyzing the delay spreads can readily be applied to the Doppler spreads. Maximum, minimum, and mean Doppler spread values are similar (within \SI{8}{\hertz} in regions \reg{1} to \reg{7}) for \BSconf{} 2 and 3, as the array aperture and position are identical. Region \reg{8}, however, shows an increased maximum Doppler spread that is caused most probably by a blocking of some \acp{RU} by a large structure when the UE moves through. 

The standard deviation is again and not surprisingly lowest for \BSconf{} 1. It is also consistently lower for the cell-free \BSconf{} 3 compared to the distributed configuration 2, a fact that has consequences on channel aging as detailed in \Cref{sec:MassiveMIMOProcessingAndResults}.
\begin{table*}\centering
	\caption{Maximum, minimum and mean value of the \ac{RMS} Doppler spread as well as its standard deviation around the mean within regions \reg{1} to \reg{8} for \BSconf{} 1 to 3.\label{tab:DopplerSpreadStats}}
	\begin{tabular}{@{}rrrrrrrrr@{}}\toprule
		& \multicolumn{8}{c}{regions}\\
		\cmidrule{2-9}
		Doppler spread $\sigma_{\indBS,\indMS}^{(\nu)}[l]$ & \reg{1} & \reg{2} & \reg{3} & \reg{4} & \reg{5} & \reg{6} & \reg{7} & \reg{8}\\ \midrule
		\ac{BS} array config. 1\\
		Max. [\SI{}{\hertz}] &  79.6&  54.0&  85.5& 109.8&  84.6&  74.1&  69.2& 125.4\\
		Min. [\SI{}{\hertz}] &  23.6&  20.7&  20.7&  53.9&  40.6&  20.6&  26.1&  43.9\\
		Mean [\SI{}{\hertz}] &  53.0&  32.9&  44.9&  87.2&  62.7&  47.6&  42.6&  84.2\\
		Std. [\SI{}{\hertz}] &   2.0&   1.3&   2.9&   4.9&   3.4&   2.4&   1.6&   3.1\\
		\ac{BS} array config. 2\\
		Max. [\SI{}{\hertz}] &  74.5&  63.3&  78.7& 113.5& 133.1& 117.3& 138.8& 116.8\\
		Min. [\SI{}{\hertz}] &  27.4&  18.5&  11.8&  41.8&  45.6&  25.3&  27.3&  28.8\\
		Mean [\SI{}{\hertz}] &  52.0&  32.6&  32.5&  76.9&  84.2&  69.2&  73.6&  75.0\\
		Std. [\SI{}{\hertz}] &   6.6&   4.0&  12.2&  11.5&  22.5&  21.0&  35.7&  20.9\\
		\ac{BS} array config. 3\\
		Max. [\SI{}{\hertz}] &  78.8&  68.7&  85.3& 116.1& 132.8& 120.4& 145.6& 131.2\\
		Min. [\SI{}{\hertz}] &  35.7&  22.1&  12.0&  38.7&  40.8&  23.7&  28.1&  29.2\\
		Mean [\SI{}{\hertz}] &  58.9&  36.2&  36.0&  80.6&  86.7&  75.5&  74.8&  82.6\\
		Std. [\SI{}{\hertz}] &   5.1&   3.7&  10.5&  10.9&  15.9&  19.0&  26.8&  20.3\\
		\bottomrule
	\end{tabular}
\end{table*}

\subsection{Collinearity of the \ac{LSF} in space}
We evaluate the collinearity of the \ac{LSF} in \cref{eqn:LFSCollinearity} to assess the stationarity of the three proposed \BSconf{} 1 to 3 in space, \ie{,} we analyze if the statistical parameters of the wireless propagation channel vary over \acp{RU}. A collinearity value close to one indicates a similar distribution of multipath components in the delay-Doppler domain, while a value close to zero indicates no similarity.

\paragraph*{Collinearity Over Distance Traveled}

\Figure[!h]()[width=\textwidth]{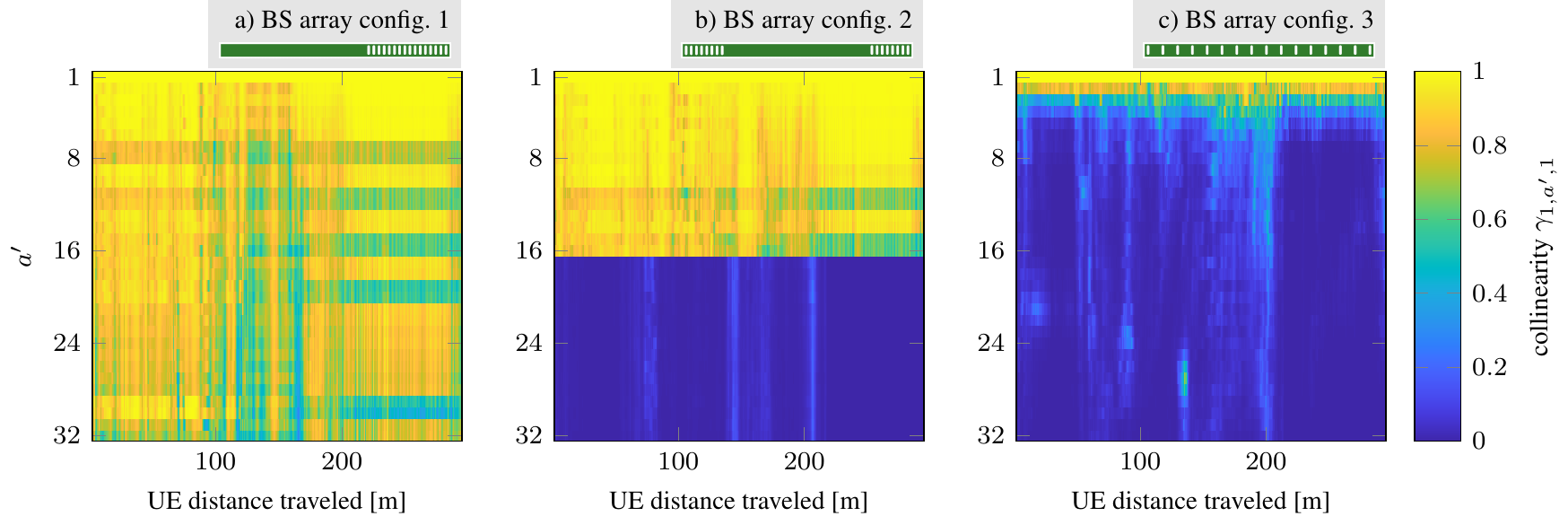}{Spatial collinearity to \ac{RU} $\indBS=1$ over distance traveled on the users trajectory for (a) \BSconf{} 1, (b) \BSconf{} 2, and (c) \BSconf{} 3. \label{fig:SpatialCollinearity_SC1-3_R2_A1toA1-32}}

\Cref{fig:SpatialCollinearity_SC1-3_R2_A1toA1-32} displays the collinearity $\gamma_{\indBS,\indBS',\indMS}[l]$ from \ac{BS} antenna $\indBS=1$ to all other \acp{RU} $\indBS'\in\{1,\ \ldots, 32\}$ over the distance traveled by $\mathrm{UE}_1$. \Cref{fig:SpatialCollinearity_SC1-3_R2_A1toA1-32} (a) for \BSconf{} 1 shows high collinearity values $\gamma_{\indBS,\indBS',\indMS}[l]$ in general ($0.8 - 1$), but also exhibits regions of lower collinearity, \eg{,} in the \ac{NLOS} regions between \SI{100}{\meter} and \SI{180}{\meter} traveled. Surprisingly, we do not see a monotonical decrease of collinearity with increasing \ac{RU} index $\indBS'$ (\ie{,} increasing distance between \acp{RU} $\indBS'$ and $\indBS$). Especially after \SI{200}{\meter} distance traveled, a repetitive pattern emerges with collinearity values of $0.9$ and $0.6$ alternating every two antennas. The reason for this is not yet clear and subject of further investigation.

\Cref{fig:SpatialCollinearity_SC1-3_R2_A1toA1-32} (b) displays the collinearity $\gamma_{\indBS,\indBS',\indMS}[l]$ for \BSconf{} 2. The top half of the plot, representing the first \ac{BS} antenna group, largely resembles the \BSconf{} 1 case. The bottom half of the plot, representing the \acp{RU} $\indBS'$ with a distance of at least \SI{45}{\meter} to antenna $\indBS=1$, does not show any significant collinearity, as expected.

\Cref{fig:SpatialCollinearity_SC1-3_R2_A1toA1-32} (c) displays the collinearity $\gamma_{\indBS,\indBS',\indMS}[l]$ for \BSconf{} 3. Already for the second \ac{BS} antenna $\indBS'=2$, collinearity drops to $0.8$ as the distance of $\SI{1.5}{\meter}=16\lambda$ causes significant changes in the propagation environment. This effect is also confirmed in \BSconf{} 1 for \acp{RU} with similar distance. For \ac{BS} antenna $\indBS'=3$ the collinearity drops further to $0.5$. We conclude that for distances larger than $\SI{3}{\meter}=32\lambda$, non-stationarity in space must be assumed. Additionally, a distance significantly smaller than that (\ie{,} $\SI{0.3}{\meter}=3.2\lambda$) is no guarantee for stationarity in a rich scattering environment, as seen in \cref{fig:SpatialCollinearity_SC1-3_R2_A1toA1-32} (a).

\paragraph*{Average Collinearity Over All Regions}

\Figure[!h]()[width=\textwidth]{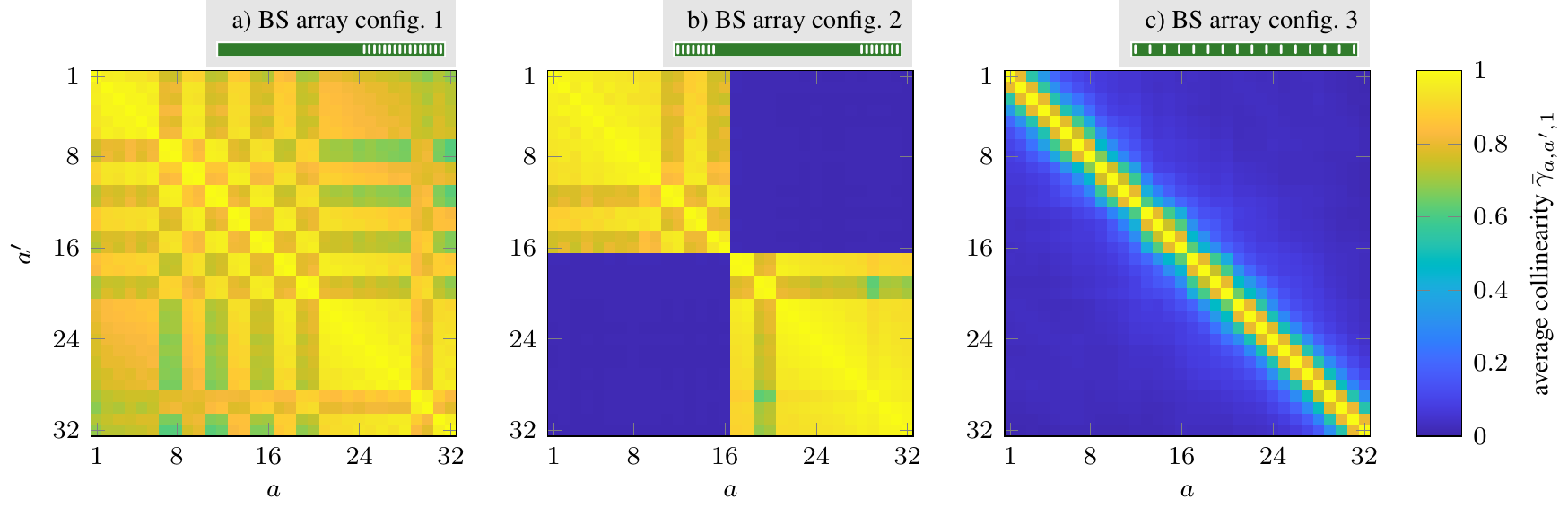}{Average spatial collinearity over all regions for all \ac{RU} combinations $(\indBS,\indBS')$ and for (a) \BSconf{} 1, (b) \BSconf{} 2, and (c) \BSconf{} 3. \label{fig:SpatialCollinearity_SC1-3_R2_A1-32toA1-32}}

We revert to the average collinearity over all regions \reg{1} to \reg{8} as defined in \cref{eqn:avrgLFSCollinearity} to analyze collinearity between all \ac{BS} antenna pairs $(\indBS,\indBS')$. This allows us to assess stationarity of antenna groups within the \ac{BS} array in space.

\Cref{fig:SpatialCollinearity_SC1-3_R2_A1toA1-32} (a) displays the average collinearity $\bar{\gamma}_{\indBS,\indBS',\indMS}[l]$ for \BSconf{} 1. Overall, collinearity of the involved \acp{LSF} is high at values between $0.9$ and $1$. However, \acp{RU} with indices $\indBS\in\{7,8,11,12,15,16,19,20,29,30\}$ apparently exhibit differing propagation characteristics as the complimentary \ac{BS} antenna set and therefore low collinearity at $0.6-0.7$, thus yielding the observed checkerboard pattern.

\Cref{fig:SpatialCollinearity_SC1-3_R2_A1toA1-32} (b) displays the average collinearity $\bar{\gamma}_{\indBS,\indBS',\indMS}[l]$ for \BSconf{} 2. The observations from above are immediately applicable to each $16\times16$ block on the main diagonal, \ie{,} the distributed \ac{BS} antenna array groups exhibit high collinearity within themselves, but not among them. Collinearity between any \ac{RU} in the east ($\indBS\in\{1,\ \ldots,16\}$) and in the west ($\indBS\in\{17,\ \ldots,32\}$) is negligible. 

\Cref{fig:SpatialCollinearity_SC1-3_R2_A1toA1-32} (c) displays the average collinearity $\bar{\gamma}_{\indBS,\indBS',\indMS}[l]$ for \BSconf{} 3. In this configuration, the distance between consecutive \acp{RU} is \SI{1.5}{\meter} and therefore the collinearity decreases rapidly, \ie{,} to $0.8$ for $|\indBS'-\indBS|=1$ and to $0.5$ for $|\indBS'-\indBS|=2$. We conclude that there is no antenna pair $(\indBS,\indBS')$ for which stationarity holds if their distance is larger than $\SI{3}{\meter}=32\lambda$.

\subsection{Key findings}
Summarizing this result section, we note the following.
\begin{itemize}
	\item Large-scale fading and shadowing cause path-loss differences of up to \SI{20}{\decibel} over horizontal linear \ac{BS} array apertures that are \SI{46.5}{\meter} in size for the scenario under investigation.
	\item The delay and Doppler spread vary greatly, up to $\SI{300}{\nano\second}$ and $\SI{100}{\hertz}$ respectively, over the aperture size for \BSconf{} 2 and 3.
	\item The standard deviation of the delay spread and the Doppler spread over \ac{RU} realizations is larger for \BSconf{} 2 than for configuration 3.
	\item Stationarity in space among \acp{RU} is likely for aperture sizes below $\SI{2}{\meter}=21\lambda$, but not guaranteed.
	\item Non-stationarity in space among \acp{RU} can be assumed for \ac{BS} antenna distances greater than $\SI{3}{\meter}=32\lambda$.
\end{itemize}

\renewcommand{\indMS}{k}
\section{Massive \ac{MIMO} processing and results}
\label{sec:MassiveMIMOProcessingAndResults}
The unmatched spectral efficiency of massive \ac{MIMO} systems \cite{Marzetta2010} arises due to $\numMS$ users being served by a large number of \acp{RU} $\numBS\gg\numMS>1$ in favorable propagation conditions within a rich scattering environment. 

A common step to simplify analysis in literature is to assume the matrix describing the wireless channel to be composed of \ac{i.i.d.} channel realizations for the \acp{RU} $\indBS$. With the growing array aperture size of distributed massive \ac{MIMO} and cell-free systems, and the resulting variation in large-scale fading, shadowing, and relative velocities over \acp{RU} as shown in the previous section, this assumption no longer holds. It is therefore not sufficiently known in the community if the linear processing algorithms commonly utilized in massive \ac{MIMO} maintain their close-to-optimal performance in cell-free systems.

In the following, we utilize the wireless channel measurements described in \Cref{sec:ScenarioDescriptionAndMeasurementFramework} and \Cref{sec:ChannelSoundingSignalModel} to derive a massive \ac{MIMO} signal model with channel matrix realizations deduced from real-world propagation scenarios. These channel matrices obtain their characteristics solely from the geometric relation between users and \ac{BS} and all the scattering and blocking objects in their surroundings as shown in \cref{fig:ScenarioOverview}. No statistical or other modeling assumptions for the channel matrix are made. 

\subsection{Massive \ac{MIMO} signal model}
\label{subsec:MassiveMIMOSignalModel}
We consider an uplink massive \ac{MIMO} system where $\numMS=2$ users $\mathrm{UE}_\indMS$ are transmitting to a \ac{BS} deploying $\numBS$ antennas. The channel vector for user $k$ at symbol index $\indDelay$ is constructed by assembling the coherently measured and sampled channel transfer function $\hat{H}_{\indBS,\indMS}[m]$ from  \cref{eqn:TFest} into vector form,  
\begin{equation}
	\HsinglesymbNoDeps{}\depsMSDelay = \left[\hat{H}_{0,\indMS}[m],\ \hat{H}_{1,\indMS}[m],\ \dots,\ \hat{H}_{\numBS-1,\indMS}[m]\right]\Transp  \in \mathbb{C}^{\numBS\times 1}.
	\label{eqn:channelVector}
\end{equation}
The channel vector collects the channel coefficients from user $\indMS$ to all $\numBS$ \acp{RU}. The frequency index $\indTones$, which is interpreted as a subcarrier index in this context, is dropped as all following analysis only considers one subcarrier at a time.

The channel vectors of all users are grouped into the channel matrix (refer to  \cite{Truong2013} for more details)
\begin{equation}
	\Hsymb{\indDelay} = \left[\Hsinglesymb{0,\indDelay},\ \Hsinglesymb{1,\indDelay},\ \dots\ \Hsinglesymb{\numMS-1,\indDelay}\right],
	\label{eqn:ChannelMatrix}
\end{equation}
where in our case $\numMS=2$, as two distinct transmitters were utilized in the channel sounding campaign.

By applying a beamforming matrix $\BF{\indDelay}= [\BFvec{0,\indDelay},\ $ $\BFvec{1,\indDelay},\ \dots\ \BFvec{\numMS-1,\indDelay}] \in \mathbb{C}^{\numBS\times\numMS}$ in the uplink, the vector collecting the received uplink symbols from all $\numMS$ users is {\cite{Zemen2019,Truong2013}}
\begin{equation}
	\tilde{\RXsymb{}}\depsDelay = \BF{\indDelay}\Herm\Hsymb{\indDelay}\TXsymbNoDeps{}\depsDelay + \frac{1}{\sqrt{P}}\tilde{\Nsymb{}}\depsDelay{,}
\end{equation}
with $P$ the average transmit power of each user, $\TXsymbNoDeps{}\depsDelay$ the vector collecting the transmit symbols of each user, and $\frac{1}{\sqrt{P}}\tilde{\Nsymb{}}\depsDelay = \frac{1}{\sqrt{P}}\BF{\indDelay}\Herm\Nsymb{}\depsDelay\DistAsComplexGauss{0,\frac{\sigma^2}{P}\eye{\numMS}}$ filtered complex Gaussian noise.

The received symbol from user $\indMS$ at the \ac{BS} then reads as {\cite{Zemen2019,Truong2013}}
\begin{equation}
	\tilde{\RXsinglesymb{}}\depsMSDelay = \BFvecNoDeps{}\depsMSDelay\Herm \HsinglesymbNoDeps{}\depsMSDelay \TXsinglesymb{}\depsMSDelay + \frac{1}{\sqrt{P}}\tilde{\NsinglesymbNoDeps{}}\depsMSDelay + \sum_{\indMS'\neq\indMS} \BFvecNoDeps{}\depsMSDelay\Herm \HsinglesymbNoDeps{}{}\depsMSpDelay\TXsinglesymb{}\depsMSpDelay{,}
	\label{eqn:SymbolSignalModel}
\end{equation}
where the first term is the desired signal, the second term is filtered and scaled Gaussian noise $\frac{1}{\sqrt{P}}\tilde{\NsinglesymbNoDeps{}}\depsMSDelay\DistAsComplexGauss{0,\frac{\sigma^2}{P}}$, and the third term is interference from other users $\indMS'\neq\indMS$.

Note that the measurement framework shown in \cref{sec:FlexibleWideApertureArrayMeasurementFramework} in itself does not introduce any inter-user interference as the users transmit their respective sounding sequence separated in time. However, interpreting the measured channel transfer function as entries in the channel matrix \cref{eqn:ChannelMatrix} provides a signal model that assumes simultaneous transmission of all users, thus yielding the interference term in \cref{eqn:SymbolSignalModel}.

%

The performance of a massive \ac{MIMO} systems in terms of \ac{SINR} (or equivalently spectral efficiency) is largely determined by the properties of the current channel matrix realization $\Hsymb{\indDelay}$ and its statistics. We therefore establish in the following the methods to assess the characteristics of the channel matrix and, ultimately, the quality of a wireless communication system based on it.  

%

\paragraph*{\ac{SINR} and Channel Aging}
Rate and reliability for user $\indMS$ are mainly determined by the instantaneous \ac{SINR} {\cite{Truong2013}}
\begin{equation}
	\SINR\depsMSDelay = \frac{\left|\BFvecNoDeps{}\depsMSDelay\Herm \HsinglesymbNoDeps{}\depsMSDelay\right|^2}{\frac{\sigma^2}{P} + \sum_{\indMS'\neq\indMS} \left|\BFvecNoDeps{}\depsMSDelay\Herm \HsinglesymbNoDeps{}{}\depsMSpDelay\right|^2}{,}
	\label{eqn:instantSINR}  
\end{equation}
defined as the ratio of the signal component to the interference and noise component in \cref{eqn:SymbolSignalModel}. We consider in what follows the beam-forming vectors $\BFvecNoDeps{}\depsMSDelay$ to being calculated via the \ac{RZF} approach {\cite{Bjornson2017,Ngo2015}} by solving
\begin{align}
	\BF{\indDelay} &= \left[\BFvecNoDeps{}_{0,\indDelay},\ \BFvecNoDeps{}_{1,\indDelay},\ \dots\ \BFvecNoDeps{}_{\numMS-1,\indDelay}\right]\\
	&= \tilde{\HsymbNoDeps}_{\indDelay,\Delta t} \left(\tilde{\HsymbNoDeps}_{\indDelay,\Delta t}\Herm\tilde{\HsymbNoDeps}_{\indDelay,\Delta t} + \frac{\sigma^2}{P^{\mathrm{(P)}}}\eye{\numMS} \right)^{-1},
	\label{eqn:BFmatrixRZF}
\end{align}
{which is approximated well by the \ac{ZF} solution 
\begin{equation}
	\BF{\indDelay} \approx \tilde{\HsymbNoDeps}_{\indDelay,\Delta t} \left(\tilde{\HsymbNoDeps}_{\indDelay,\Delta t}\Herm\tilde{\HsymbNoDeps}_{\indDelay,\Delta t}\right)^{-1},
	\label{eqn:BFmatrix}
\end{equation}
if $P^{\mathrm{(P)}}\gg \sigma^2$, \ie, if the pilot power during \ac{CSI} acquisition is far greater than the noise power. Since the measurement \ac{SNR} for obtaining the channel vector realizations in \cref{eqn:channelVector} is greater than \SI{25}{\decibel} at all times, $\tilde{\HsymbNoDeps}_{\indDelay,\Delta t}\Herm\tilde{\HsymbNoDeps}_{\indDelay,\Delta t} \gg \frac{\sigma^2}{P^{\mathrm{(P)}}}\eye{\numMS}$ holds and the regularization term is therefore omitted when evaluating the results in what follows. Simulation results (not shown) confirm this approach, as the maximum difference in median \ac{SINR} utilizing either \ac{RZF} beam-forming in \cref{eqn:BFmatrixRZF} or \ac{ZF} beam-forming in \cref{eqn:BFmatrix} is found to be below \SI{0.3}{\decibel} throughout all simulated scenarios.}


To quantify the effect of channel aging, we deliberately introduce outdated channel matrices to calculate the beam-forming matrices in \cref{eqn:BFmatrix}. The outdated channel matrix $\tilde{\HsymbNoDeps}_{\indDelay,\Delta t}$ is defined similar to \cref{eqn:ChannelMatrix} as the outdated channel vectors $\tilde{\HsinglesymbNoDeps{a}}_{\indMS,\indDelay,\Delta t}$ grouped into a matrix, \ie{,}
\begin{align}
	\tilde{\HsymbNoDeps}_{\indDelay,\Delta t} = &\left[\tilde{\HsinglesymbNoDeps{a}}_{0,\indDelay,\Delta t},\ \tilde{\HsinglesymbNoDeps{a}}_{1,\indDelay,\Delta t},\ \dots\ \tilde{\HsinglesymbNoDeps{a}}_{\numMS-1,\indDelay,\Delta t}\right],\\
	\begin{split}
		\tilde{\HsinglesymbNoDeps{a}}_{\indMS,\indDelay,\Delta t} = &\Big[\hat{H}_{0,\indMS}(mT_\mathrm{R}+\Delta t),\ \hat{H}_{1,\indMS}(mT_\mathrm{R}+\Delta t),\\
		& \qquad \dots\ ,\ \hat{H}_{\numBS-1,\indMS}(mT_\mathrm{R}+\Delta t)\Big]\Transp.
	\end{split}
\end{align} 

The channel transfer functions $\hat{H}_{\indBS,\indMS}(mT_\mathrm{R}+\Delta t)$ that were not measured directly are obtained by interpolating the preceding and subsequent measured channel transfer functions $\hat{H}_{\indBS,\indMS}[m'] = \hat{H}_{\indBS,\indMS}(m'T_\mathrm{R}),\ \indDelay'\in\{\indDelay-2,\indDelay-1,\indDelay,\indDelay+1\}$ using a cubic spline interpolation \cite{mckinley1998cubic,de1978practical}. This approach works sufficiently well since the repetition rate of the channel sounding measurements $T_\mathrm{R}=\SI{1}{\milli\second}$ is high enough to periodically capture channel transfer function realizations while the users move maximally $\SI{1.7}{\centi\meter} = 0.18\lambda$ at their maximum velocity of \SI{60}{\kilo\meter\per\hour} \cite{Coleri2002}.

The beam-forming matrix $\BF{\indDelay}$ determines the ability of a massive \ac{MIMO} system to maximize the signal component and minimize interference from other users. Acquisition of timely \ac{CSI} (\ie{,} with small delay $\Delta t$) $\tilde{\HsymbNoDeps}_{\indDelay,\Delta t}\approx{\HsymbNoDeps}_{\indDelay}$ that is necessary to calculate the beam-forming matrix is, however, non-trivial if high mobility is involved. By the time the beam-forming matrix is applied, it might already be out-dated (\ie{,} $\tilde{\HsymbNoDeps}_{\indDelay,\Delta t}\not\approx{\HsymbNoDeps}_{\indDelay}$) and the instantaneous \ac{SINR} in \cref{eqn:instantSINR} decreases. This effect is called \textit{channel aging} and prevents massive \ac{MIMO} systems to operate in high mobility scenarios without special measures such as channel prediction \cite{Loeschenbrand2020}.

\oldnew{\Ac{MMSE} beam-forming is not considered because its benefits over \ac{ZF} are negligible in the high \ac{SNR} regime investigated in this work (\ie \ac{CSI} is acquired during a channel sounding campaign where the \ac{UE} transmit power is \SI{38}{\deci\belm} and the measurement \ac{SNR} is greater than \SI{25}{\decibel} at all times).}{}

\paragraph*{Channel Hardening} 
The large number of \acp{RU} $\numBS \gg \numMS \gg 1$ leads to linear beam-forming (\eg{,} \ac{ZF} as outlined above) being close to optimal. Additionally, the law of large numbers guarantees that random fluctuations of the signal component $|\BFvecNoDeps{}\depsMSDelay\Herm \HsinglesymbNoDeps{}\depsMSDelay|^2$ become less probable and the effective channel $\BFvecNoDeps{}\depsMSDelay\Herm \HsinglesymbNoDeps{}\depsMSDelay$ becomes quasi-deterministic -- a process called \textit{channel hardening} \cite{Zemen2019}. 

As a measure for channel hardening, we revert to the signal component's ratio of the standard deviation estimation to its estimated mean over $\numDelay$ consecutive time indices {\cite{Zemen2019}}
\begin{align}
	\gamma_{\indMS,l} &= \frac{\sqrt{\dfrac{1}{\numDelay-1}\displaystyle\sum_{\indDelay=-\numDelay/2+l\numDelay}^{\numDelay/2-1+l\numDelay} \left(\left|\BFvecNoDeps{}\depsMSDelay\Herm \HsinglesymbNoDeps{}\depsMSDelay\right|^2-\mu_{\indMS,l}\right)^2}} {\mu_{\indMS,l}}\label{eqn:ChannelHardening}{,}\\
	\mu_{\indMS,l} &= \frac{1}{\numDelay}\sum_{\indDelay=-\numDelay/2+l\numDelay}^{\numDelay/2-1+l\numDelay} \left|\BFvecNoDeps{}\depsMSDelay\Herm \HsinglesymbNoDeps{}\depsMSDelay\right|^2{,}
\end{align}
which tends to zero as the channel becomes more and more deterministic.

\subsection{Massive \ac{MIMO} Results}
This section presents the results for the instantaneous \ac{SINR}, channel aging and channel hardening derived from the measurement results obtained as described in \cref{sec:ScenarioDescriptionAndMeasurementFramework} and the signal model introduced in \cref{subsec:MassiveMIMOSignalModel}. The measured channel transfer function realizations are directly interpreted as channel vector realizations as defined in \cref{eqn:channelVector}. {Since the number of supported \ac{BS} antennas is limited to 32 by the measurement framework, no analysis with an increased number of antennas is presented.}

\paragraph*{\ac{SINR} and Channel Aging Results}
The instantaneous \ac{SINR} \cref{eqn:instantSINR} determines achievable rates for the respective user $\indMS$ at a given time instant $\indDelay$. The signal component in the numerator and the interference component in the denominator heavily depend on the beam-forming applied at the \ac{BS}, which in turn relies on timely \ac{CSI}. We analyze the influence of aged \ac{CSI} on the instantaneous \ac{SINR} by calculating the beam-forming matrix \cref{eqn:BFmatrix} with outdated channel matrices $\tilde{\HsymbNoDeps}_{\indDelay,\Delta t}$ that are delayed by $\Delta t \in \{\SI{10}{\micro\second}, \SI{100}{\micro\second},\SI{500}{\micro\second}\}$. {The average noise power to transmit power ratio $\frac{\sigma^2}{P}$ in \cref{eqn:instantSINR} is chosen to be \SI{-112}{dB} in all simulations. With an average path-loss of \SI{-92}{\decibel} in \BSconf{} 1 over all regions \reg{1} to \reg{8}, the average over the time-varying \ac{SNR} in all regions without interference is \SI{20}{\decibel}.}

\Figure[!h]()[width=\textwidth]{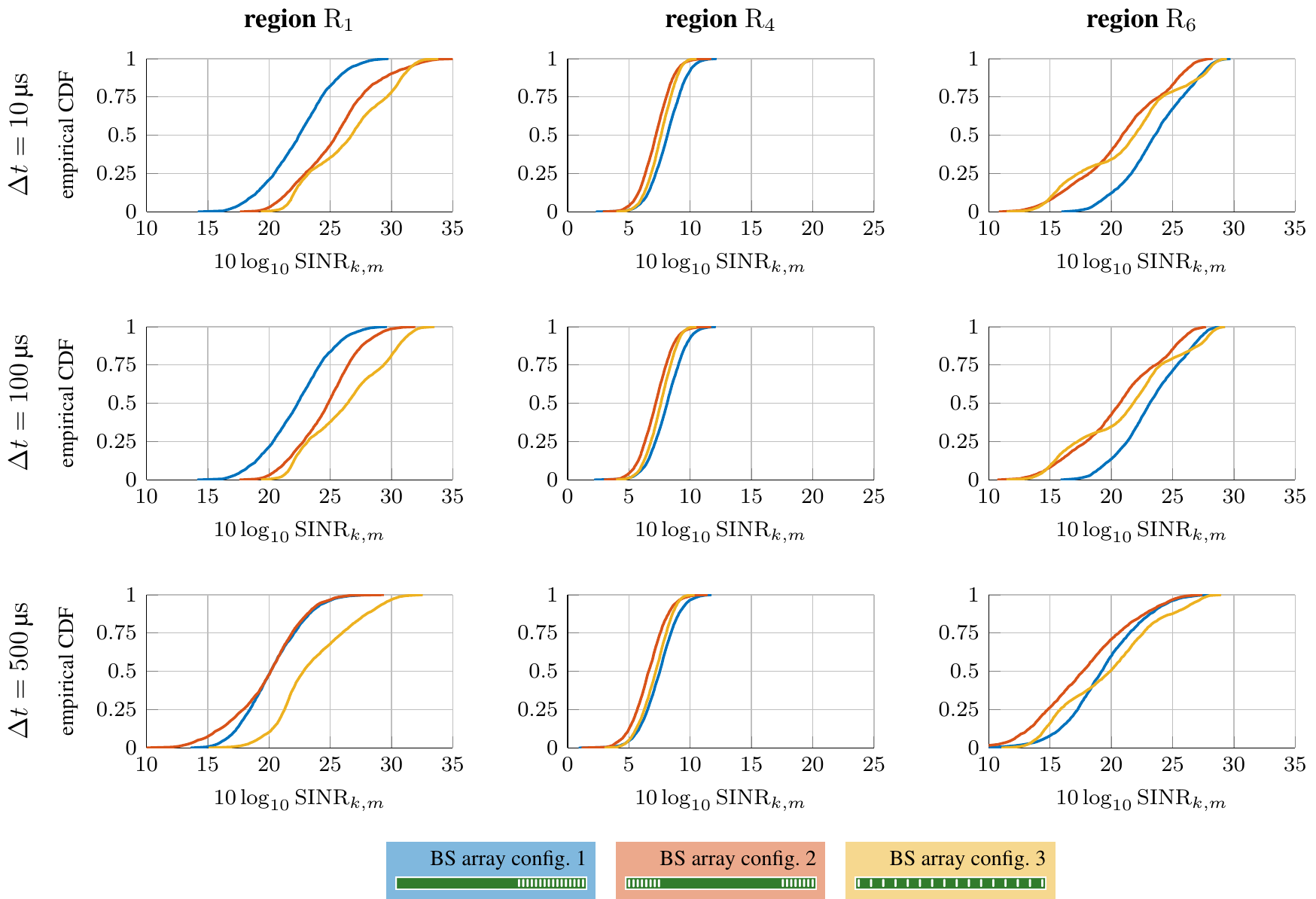}{Empirical \ac{CDF} of the \ac{SINR} for regions \reg{1} (left), \reg{4} (middle) and \reg{6} (right), \BSconf{} 1 to 3, and channel matrices $\tilde{\HsymbNoDeps}_{\indDelay,\Delta t}$ aged $\Delta t = \SI{10}{\micro\second}$ (top), $\Delta t = \SI{100}{\micro\second}$ (middle), and $\Delta t = \SI{500}{\micro\second}$ (bottom). \label{fig:ECDF_SINR_SC123_R2_age10-100-500us_reg146}}

\Cref{fig:ECDF_SINR_SC123_R2_age10-100-500us_reg146} plots the empirical \ac{CDF} of the \ac{SINR} \cref{eqn:instantSINR} for outdated channel matrices (vertically aligned) and different regions (horizontally aligned). To calculate the empirical \ac{CDF}, all channel matrices obtained in a given region (in which the \acp{UE} move along their trajectory) are used to calculate \cref{eqn:BFmatrix} and in turn the \ac{SINR}. Each subplot in \cref{fig:ECDF_SINR_SC123_R2_age10-100-500us_reg146} shows the three \BSconf{} 1 to 3 under consideration in this paper. Regions \reg{1} and \reg{6} exhibit \ac{LOS} propagation characteristics while region \reg{4} is purely \ac{NLOS}, see \cref{fig:ScenarioOverview}. 

In region \reg{1} and with small channel aging of $\Delta t = \SI{10}{\micro\second}$, \BSconf{} 3 shows the highest \ac{SINR} values on average, and \BSconf{} 2 performs slightly worse. \BSconf{} 1, however, has a significant drawback since the average distance between \acp{RU} and \acp{UE} is highest in this configuration (the larger aperture of configuration 2 and 3 lead to \acp{RU} being closer to the \ac{UE} trajectory). The \ac{SINR} is on average \SI{3}{\decibel} lower in this configuration.

When the delay between \ac{CSI} acquisition and beam-forming increases to $\Delta t = \SI{100}{\micro\second}$, the \ac{BS} array configuration most affected is \BSconf{} 2 which now only shows slightly higher average \ac{SINR} values than configuration 1. \Ac{BS} configuration 1 and 3 largely maintain the distribution of the empirical \ac{CDF}. For all three configurations, the effect of channel aging is already noticeable

For a delay between \ac{CSI} acquisition and beam-forming of $\Delta t = \SI{500}{\micro\second}$, as depicted in \cref{fig:ECDF_SINR_SC123_R2_age10-100-500us_reg146} at the bottom left, the \ac{SINR} drops significantly for all \BSconf{} 1 to 3. {Specifically, the median \ac{SINR} value over the region \reg{1} drops by \SI{4}{\decibel} in configuration 1, \SI{8}{\decibel} in configuration 2, and \SI{6}{\decibel} in configuration 3, compared to hardly any channel aging (at a \ac{CSI} delay of \SI{10}{\micro\second}).} 

The comparatively small effects of channel aging in region \reg{1} on \BSconf{} 1 is explained by the small relative velocity of the \acp{UE} in relation to the \ac{BS}. Both configuration 2 and 3 feature \acp{RU} which are closer to the \ac{UE} trajectory, but therefore also exhibit larger Doppler spreads, see also \cref{fig:DopplerSpreadColor_SC2_R2_A1-32}.

In region \reg{4} with \ac{NLOS} propagation conditions, depicted in the center column of \cref{fig:ECDF_SINR_SC123_R2_age10-100-500us_reg146}, channel aging has only very limited effect on the empirical \ac{CDF} of the \ac{SINR}. {The median drops by at most \SI{1}{\decibel} for all \ac{BS} array configurations, comparing $\tilde{\HsymbNoDeps}_{\indDelay, \Delta t}$ aged \SI{10}{\micro\second} and \SI{500}{\micro\second}, respectively.} Again, \BSconf{} 3 shows the least sensitivity to channel aging.

In region \reg{6}, the right-most column in \cref{fig:ECDF_SINR_SC123_R2_age10-100-500us_reg146}, \BSconf{} 1 shows the clear advantage of having all \acp{RU} close to the current position of the \acp{UE} and in direct \ac{LOS}, whereas for both configuration 2 and 3 a considerable amount of antennas are still blocked by a large office building. Therefore, both large aperture configurations 2 and 3 give a similar \ac{SINR} distribution that is in general shifted to lower values when compared to \BSconf{} 1. {However, also in region \reg{6} \BSconf{} 3 is less influenced by channel aging of $\Delta t = \SI{500}{\micro\second}$, with its median \ac{SINR} value dropping only \SI{5}{\decibel} as compared to \SI{8}{\decibel} and \SI{6}{\decibel} of configuration 1 and 2, respectively.}

\paragraph*{Channel Hardening Results}
An evaluation of the channel hardening coefficient \cref{eqn:ChannelHardening} distribution in regions \reg{1} (\ac{LOS}), \reg{4} (\ac{NLOS}), and \reg{6} (\ac{LOS}) is shown in \cref{fig:channelhardening_SC123_R2_reg146}. It is immediately evident that channel hardening, \ie{,} the random fluctuations of the signal component $|\BFvecNoDeps{}\depsMSDelay\Herm \HsinglesymbNoDeps{}\depsMSDelay|^2$ in \cref{eqn:instantSINR}, is lowest for the \BSconf{} 3 in all regions. This leads to the conclusion that if deterministic signal levels are desired, cell-free configurations improve on the channel hardening capabilities of conventional massive \ac{MIMO} systems.

\Figure[!h]()[width=\textwidth]{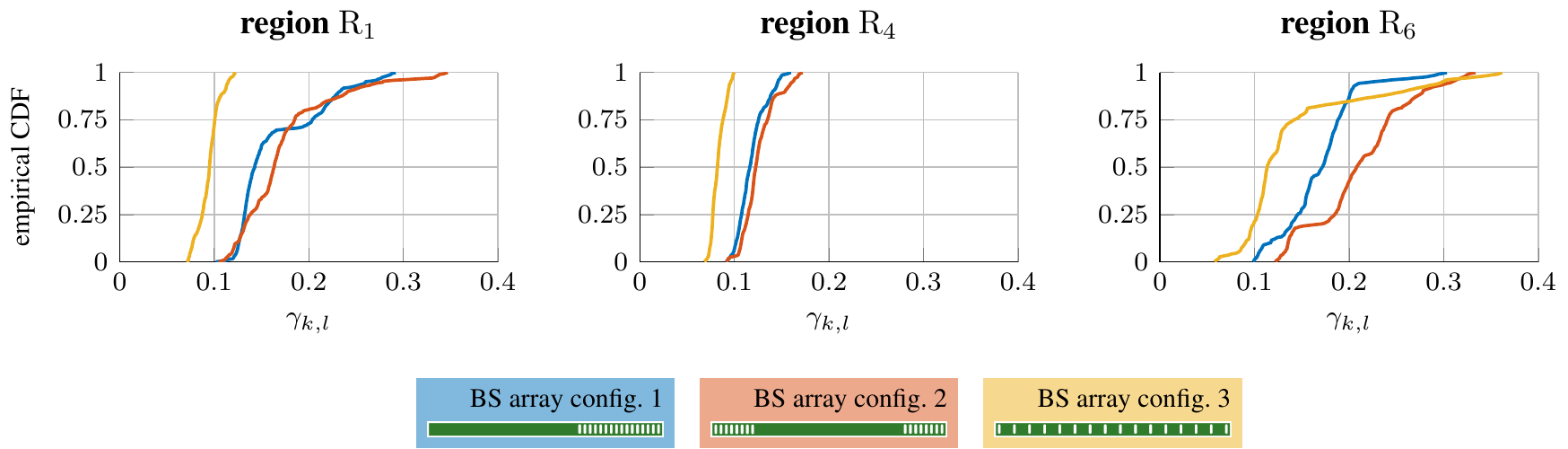}{Empirical \ac{CDF} of the channel hardening coefficient $\gamma_{\indMS,l}$ for regions \reg{1}, \reg{4} and \reg{6}, and \BSconf{} 1 to 3. \label{fig:channelhardening_SC123_R2_reg146}}

\subsection{Key Findings}
Summarizing the massive \ac{MIMO} processing result section, we note the following.
\begin{itemize}
	\item Cell-free systems are less susceptible to channel aging than distributed or conventional massive \ac{MIMO} systems as presented in \cref{fig:ECDF_SINR_SC123_R2_age10-100-500us_reg146}. The exact reason behind this phenomenon is still under investigation, but we suspect that the wide range of relative velocities of the \ac{UE} in relation to the \acp{RU} (see the Doppler spreads in \cref{fig:DelaySpreadColor_SC3_R2_A1-32}) mitigates channel aging to some extent.
	\item Cell-free systems offer better channel hardening capabilities than distributed or conventional massive \ac{MIMO} systems as presented in \cref{fig:channelhardening_SC123_R2_reg146} due to the great variation in multipath distribution over \acp{RU} (see also \cref{fig:SpatialCollinearity_SC1-3_R2_A1-32toA1-32}).
\end{itemize}

\section{Conclusion}
\label{sec:Conclusion}
In this paper we investigated radio wave propagation conditions of cell-free widely distributed massive \ac{MIMO} systems to confirm the merits promised by signal processing theory. In such systems radio units are spread out over a large geographical region and the radio signal of a \ac{UE} is coherently detected by a subset of \acp{RU} in the vicinity of the \ac{UE} and processed jointly at the nearest \ac{BPU}. Cell-free systems promise two orders of magnitude less transmit power, spatial focusing at the \ac{UE} position for high reliability, and consistent throughput over the coverage area. However, these properties have been investigated only from a theoretical point of view so far. 

We presented a \ac{SDR} based measurement system and an analysis of empirical radio wave propagation measurements in the form of time-variant channel transfer functions for a linear widely distributed antenna array with 32 single antenna \acp{RU} spread out over a range of \SI{46.5}{\meter}. Three different co-located and widely distributed \ac{RU} configurations and their properties in an urban environment have been analyzed in terms of time-variant delay-spread, Doppler spread, path-loss and the correlation of the \ac{LSF} over space. 

The path-loss from the \ac{UE} over all \acp{RU} shows a maximal variation of \SI{5}{\decibel} for a closely spaced linear array ($0.64\lambda$ spacing, \BSconf{} 1) and a variation of \SI{20}{\decibel} for the widely distributed massive \ac{MIMO} array (16$\lambda$ spacing, \BSconf{} 3). The strongest difference can be observed in the transition phase form \ac{LOS} to \ac{NLOS}.

The variance of the \ac{RMS} delay and \ac{RMS} Doppler-spread is smallest for the co-located \BSconf{} 1 and increases with the aperture of the antenna arrays. Hence, the widely distributed \BSconf{} 2 and 3 exhibit an increased range of \ac{RMS} delay spread (\SI{300}{\nano\second}) and Doppler spread (\SI{100}{\hertz}) over all 32 \acp{RU}. Also here the strongest variation is caused by the transition from \ac{NLOS} to \ac{LOS}.

The stationarity in space among \acp{RU} was validated to be likely for an aperture size smaller than $\SI{2}{\meter}=21\lambda$, although it is not guaranteed to hold. For apertures larger than $\SI{3}{\meter}=32\lambda$, non-stationary properties have been confirmed by means of the collinearity of the \ac{LSF} in space.

For the development of 6G cell-free massive \ac{MIMO} transceiver algorithms, we analyzed properties such as channel hardening, channel aging and its influence on the \ac{SINR}. Channel aging shows the strongest impact in \ac{LOS} regions for all three measured configurations. \BSconf{} 3 exhibits the best \ac{SINR} for long aging intervals in both \ac{LOS} and \ac{NLOS} conditions, even with less received power. For channel hardening, the widely-distributed \BSconf{} 3 provides the strongest effect compared to \BSconf{} 1 and 2 for \ac{LOS} as well as \ac{NLOS} scenarios. 

Our empirical evidence, summarized above, supports the promising claims for widely distributed user-centric cell-free systems as a revolutionary new 6G architecture.



\ifCLASSOPTIONcaptionsoff
  \newpage
\fi



%
\bibliographystyle{IEEEtran}
\bibliography{./bibtex/bib/IEEEabrv,./bibtex/bib/library,./bibtex/bib/IEEEexample,./bibtex/bib/CoAuthorsLiterature}

%
%

%

\begin{IEEEbiography}[{\includegraphics[width=1in,height=1.25in,clip,keepaspectratio]{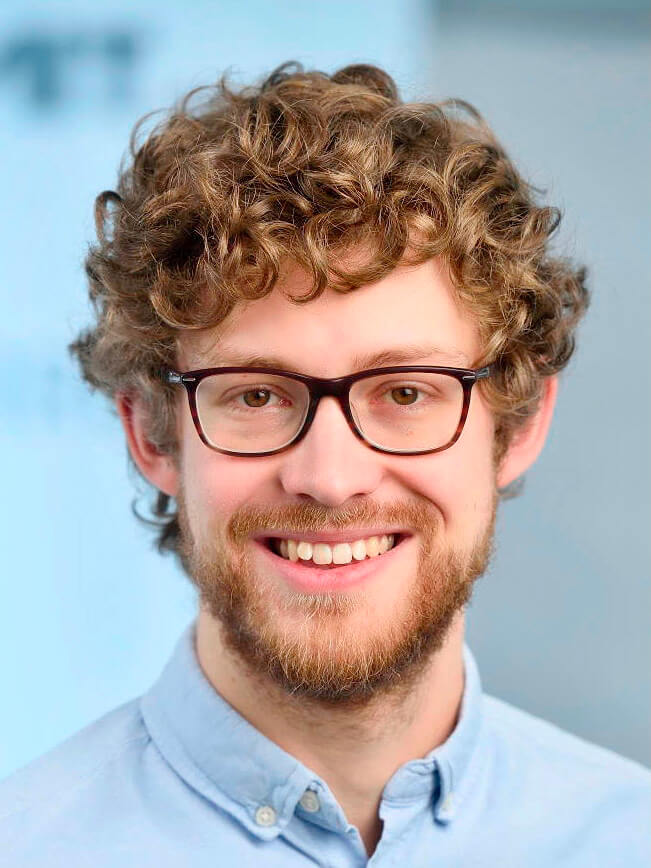}}]{David Löschenbrand} received the Dipl.-Ing. degree (with distinction) in Telecommunications in 2016 from Vienna University of Technology. From 2012 to 2015, he worked for the Institute of Telecommunications, implementing software for antenna characterization purposes. Since 2016, he is a Ph.D. candidate with the AIT Austrian Institute of Technology in Vienna, Austria. His research focuses on massive MIMO in time-varying propagation channels, channel aging, antenna design, reliable low-latency wireless communications for highly autonomous vehicles, vehicular channel measurements and channel modeling. 
\end{IEEEbiography}

\begin{IEEEbiography}[{\includegraphics[width=1in,height=1.25in,clip,keepaspectratio]{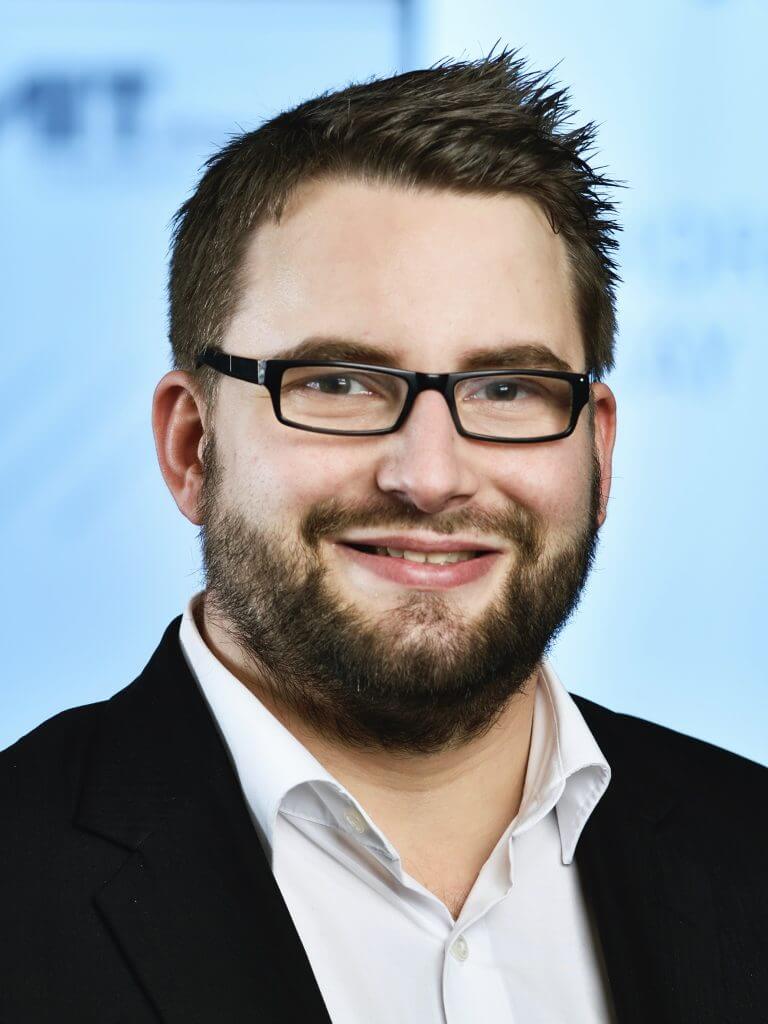}}]{Markus Hofer} received the Dipl.-Ing. degree (with distinction) in telecommunications from the Vienna University of Technology, Vienna, Austria, in 2013 and the doctoral degree in 2019. From 2013 to 2015 he was with the FTW Telecommunications Research Center Vienna working as a Researcher in the Signal and Information Processing department. He has been with the AIT Austrian Institute of Technology, Vienna since 2015 and is working as a Scientist in the research group for ultrareliable wireless machine-to-machine communications. His research interests include ultra-reliable low latency wireless communications, reflective intelligent surfaces, mmWave communications, cell-free massive MIMO, time-variant channel measurements, modeling and realtime emulation; time-variant channel estimation, 5G massive MIMO systems; software-defined radio rapid prototyping, cooperative communication systems, and interference management.
\end{IEEEbiography}

\begin{IEEEbiography}[{\includegraphics[width=1in,height=1.25in,clip,keepaspectratio]{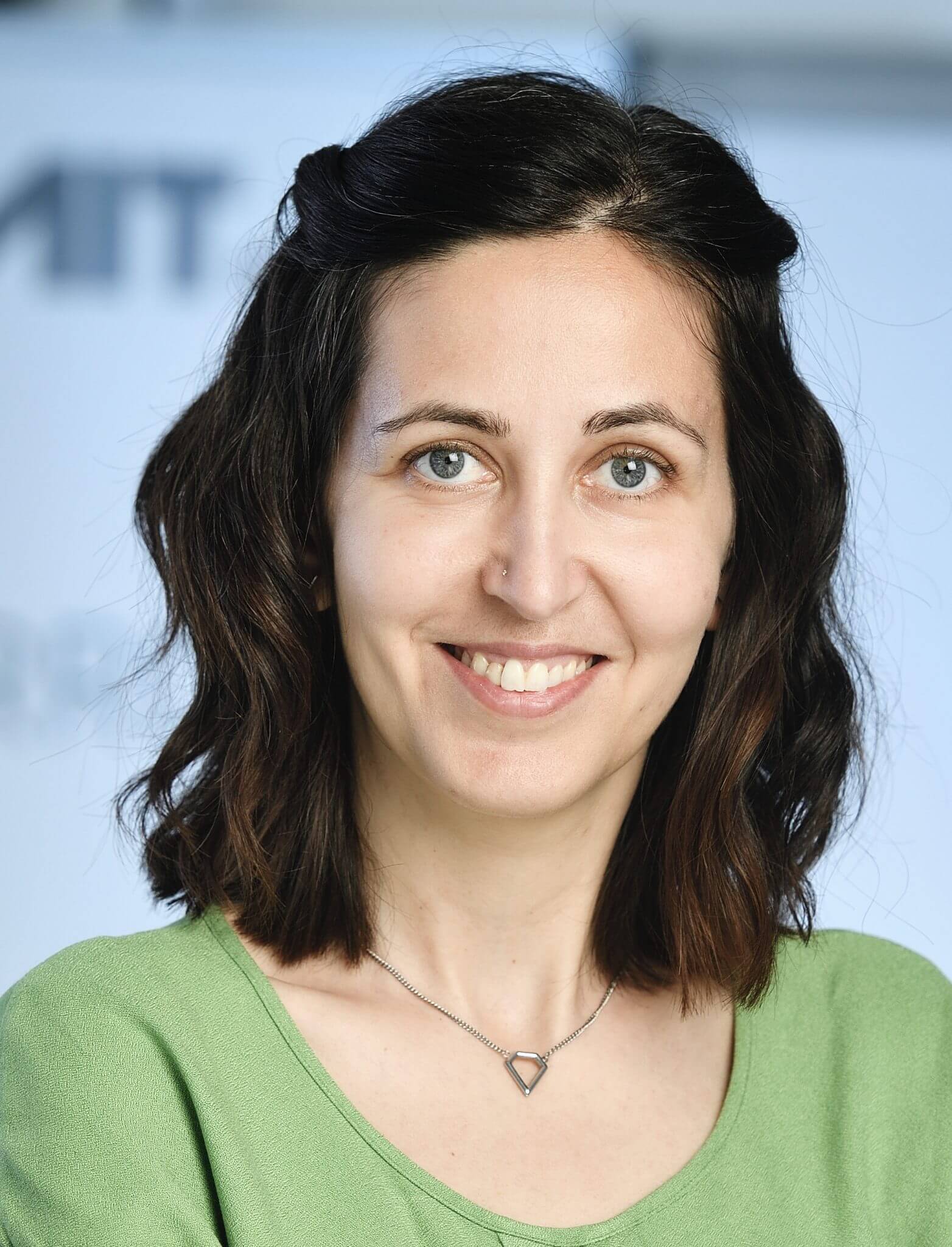}}]{Laura Bernadó} obtained her PhD in telecommunications engineering from the Vienna University of Technology (VUT) in 2012 and the M.Sc. degree in telecommunications engineering from the Technical University of Catalonia (UPC) in 2007, with the Master Thesis written at the Royal Institute of Technology (KTH), in Stockholm. Mrs. Bernadó has worked as a researcher in the signal and information processing department at the Telecommunications Research Center in Vienna, Austria, for 6 years, and as an antenna engineer at Fractus SA, Spain, for 3 years. Currently she works a scientist in the reliable wireless communications research group at the department for digital safety and security at AIT Austrian Institute of Technology. Her research interests are modeling of fast time-varying non-stationary fading processes, channel emulation and transceiver design for ultra-reliable wireless communication systems. 
\end{IEEEbiography}

\begin{IEEEbiography}[{\includegraphics[width=1in,height=1.25in,clip,keepaspectratio]{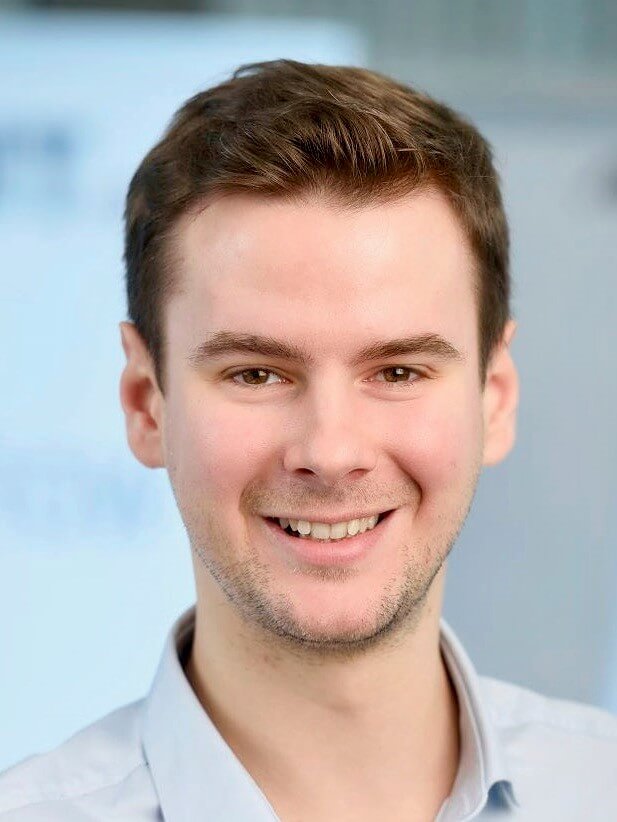}}]{Stefan Zelenbaba} graduated from the Faculty of Electrical Engineering of University of Belgrade, Telecommunications Engineering in 2015, and received his Master’s degree in collaboration with Nokia Bell Labs in 2017. Since 2017 he is a Ph.D. candidate with the Austrian Institute of Technology in the reliable wireless communications research group of Thomas Zemen. His research is focused on measurements and characterization of non-stationary time-variant wireless channels as well as geometry-based wireless channel models and their validation.
\end{IEEEbiography}

\begin{IEEEbiography}[{\includegraphics[width=1in,height=1.25in,clip,keepaspectratio]{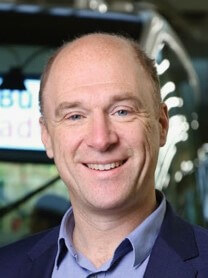}}]{Thomas Zemen} (S’03-M’05-SM’10) received the Dipl.-Ing. degree in electrical engineering in 1998, the doctoral degree in 2004 and the Venia Docendi (Habilitation) in 2013, all from Vienna University of Technology. He is Principal Scientist at the AIT Austrian Institute of Technology, Vienna, Austria, leading the reliable wireless communications group. Previously he worked at the Telecommunication Research Center Vienna (FTW) and Siemens Austria.
Mr. Zemen is the author or coauthor of four books chapters, 37 journal papers and more than 113 conference communications. His research interests focus on the interaction of the physical radio communication channel with other parts of a communication system for time-sensitive 5G and 6G applications.
Dr. Zemen is docent at the Vienna University of Technology and served as Editor for the IEEE Transactions on Wireless Communications from 2011 - 2017
	
\end{IEEEbiography}




\EOD

\end{document}